\documentclass[12pt]{article}
\usepackage{amsmath,amssymb,amsthm,amsxtra,bbm,overpic,bm,epsfig,subfigure,tikz,tikz-feynman}
\usepackage{hyperref}
\usepackage{mathrsfs}
\usepackage{enumitem}
\usepackage{graphicx}
\usepackage{color}
\usepackage{comment}
\usepackage{epstopdf}
\usepackage{float}
\usepackage{cite}
\allowdisplaybreaks[2]
\usetikzlibrary{arrows.meta}
\makeatletter
\tikzfeynmanset{compat=1.1.0} % 启用兼容模式（适用于 pdflatex）
\makeatother
\usepackage[all]{xy}

\usepackage{slashed,stmaryrd,multirow}
\numberwithin{equation}{section}

\def\thefootnote{\fnsymbol{footnote}}

\usepackage{hyperref}
\usepackage{slashed,stmaryrd,orcidlink}

\usepackage{lscape}%
\usepackage{array}
\usepackage{booktabs}%

\begin{document}
	
	\vspace{0.2cm}
	
	\begin{center}
		{\Large\bf One-loop Renormalization of the Type-I Seesaw Model \\in the Modified Minimal-subtraction Scheme}
	\end{center}
	
	\vspace{0.2cm}
	
	\begin{center}
		{\bf Jihong Huang}~{\orcidlink{0000-0002-5092-7002}},$^{1,2}$~\footnote{E-mail: huangjh@ihep.ac.cn}
		\quad
		{\bf Shun Zhou}~{\orcidlink{0000-0003-4572-9666}}~$^{1,2,3}$~\footnote{E-mail: zhoush@ihep.ac.cn (corresponding author)}
		\\
		\vspace{0.2cm}
		{\small
			$^{1}$Institute of High Energy Physics, Chinese Academy of Sciences, Beijing 100049, China\\
			$^{2}$School of Physical Sciences, University of Chinese Academy of Sciences, Beijing 100049, China\\
$^{3}$Center for High Energy Physics, Henan Academy of Sciences, Zhengzhou 450046, China}
	\end{center}

	\vspace{0.5cm}
	
\begin{abstract}
Extending the Standard Model (SM) with three right-handed neutrinos, the type-I seesaw model serves as the simplest and most natural scenario to successfully explain both tiny neutrino masses and the baryon number asymmetry in the Universe. In this paper, we perform a complete one-loop renormalization of the type-I seesaw model in the modified minimal-subtraction ($\overline{\rm MS}$) scheme. The one-loop self-energy corrections of charged leptons and Majorana neutrinos are calculated in the $R_\xi^{}$ gauge, and the explicit expressions of all the counterterms for wave functions, fermion masses and the leptonic flavor mixing matrix are given. Furthermore, adopting the Euler-like parametrization of the $6\times 6$ unitary leptonic flavor mixing matrix, we derive one-loop renormalization-group equations for all the physical parameters in the $\overline{\rm MS}$ scheme, including neutrino masses, mixing angles and CP-violating phases. The modification of the one-loop renormalization of the original SM parameters due to the presence of heavy Majorana neutrinos is investigated as well. In this way, we provide a self-consistent theoretical framework to thoroughly test the type-I seesaw model at the one-loop level with future precision data.
\end{abstract}

	\def\thefootnote{\arabic{footnote}}
	\setcounter{footnote}{0}
	
	\newpage

\section{Introduction}
	
Although a number of elegant neutrino oscillation experiments have revealed that neutrinos are massive, it remains unknown how their tiny masses are generated~\cite{ParticleDataGroup:2024cfk,Xing:2020ijf}. Among various models of neutrino mass generation, the type-I seesaw model~\cite{Minkowski:1977sc,Yanagida:1979as,Gell-Mann:1979vob,Glashow:1979nm,Mohapatra:1979ia} appears to be the simplest and most natural in the sense that only three right-handed neutrino singlets $N_{\rm R}^{}$ are introduced and all the terms in the full Lagrangian are allowed by the symmetries of the minimal Standard Model (SM). More explicitly, the gauge-invariant Lagrangian of the type-I seesaw model reads
\begin{eqnarray}\label{eq:L_SS}
{\cal L} = {\cal L}_{\rm SM}^{} + \overline{N_{\rm R}^{}} {\rm i} \slashed{\partial} N_{\rm R}^{} - \left(\overline{\ell_{\rm L}^{}} {\bf y}_\nu^{} \widetilde{H} N_{\rm R}^{} + \frac{1}{2} \overline{N_{\rm R}^{\rm c}} {\bf m}_{\rm R}^{} N_{\rm R}^{} + {\rm h.c.}\right) \;, 
\end{eqnarray}
where ${\cal L}_{\rm SM}^{}$ is the SM Lagrangian, $\widetilde{H} \equiv {\rm i} \sigma^2 H^*$ with $H$ being the SM Higgs doublet, $\ell_{\rm L}^{} = \left(\nu_{\rm L}^{}, l_{\rm L}^{}\right)^{\rm T}$ stands for the left-handed lepton doublet. In Eq.~(\ref{eq:L_SS}), the charge-conjugate counterpart of the right-handed neutrino singlet is defined as $N_{\rm R}^{\rm c} \equiv {\sf C} \overline{N_{\rm R}^{}}^{\rm T}$ with ${\sf C}$ being the charge-conjugation matrix. In addition, the Dirac neutrino Yukawa coupling matrix ${\bf y}_\nu^{}$ is in general an arbitrary $3\times 3$ complex matrix, whereas the Majorana neutrino mass matrix ${\bf m}_{\rm R}^{}$ must be symmetric.\footnote{For clarity, all the $3\times 3$ matrices in the flavor space of fermions will always be denoted by Latin letters in boldface throughout this work.} 

Depending on whether or not the absolute scale $\Lambda^{}_{\rm R}$ of the Majorana mass matrix ${\bf m}_{\rm R}^{}$ is much higher than the electroweak scale $\Lambda^{}_{\rm EW} \sim 10^2~{\rm GeV}$, one can explore the low-energy phenomenology of the type-I seesaw model in two different manners.
\begin{itemize}
    \item In the case where $\Lambda^{}_{\rm R}$ is comparable to or not far above $\Lambda^{}_{\rm EW}$, we regard the type-I seesaw model as an ultra-violet (UV) complete theory for elementary particles, including massive neutrinos as already observed in neutrino oscillation experiments. In a similar way to the establishment of the SM, one ultimately must go beyond the tree level and take into account radiative corrections to physical observables, in order to test the type-I seesaw model with precision data. As usual, we first find out the true vacuum state and convert into the physical basis, where elementary particles have definite masses and electric charges. This is achieved by requiring the Higgs field to take its vacuum expectation value, i.e., $\left<H\right> = (0,v^{}_0/\sqrt{2})^{\rm T}$ with $v^{}_0 \approx246~{\rm GeV}$. Thus the gauge symmetry ${\rm SU}(2)^{}_{\rm L}\times {\rm U}(1)^{}_{\rm Y}$ is spontaneously broken down to ${\rm U}(1)^{}_{\rm em}$ and we obtain the Dirac neutrino mass matrix ${\bf m}_{\rm D}^{} = {\bf y}_\nu^{} v^{}_0 /\sqrt{2}$. The overall neutrino mass term becomes 
	\begin{eqnarray}
		\label{eq:mass_term}
		{\cal L}_{\rm mass}^{} = -\frac{1}{2} \overline{\begin{pmatrix} \nu_{\rm L}^{} & N_{\rm R}^{\rm c} \end{pmatrix}} 
		\begin{pmatrix}
			{\bf 0} & {\bf m}_{\rm D}^{} \\ {\bf m}_{\rm D}^{\rm T} & {\bf m}_{\rm R}^{} 
		\end{pmatrix} 
		\begin{pmatrix}
			\nu_{\rm L}^{\rm c} \\ N_{\rm R}^{} 
		\end{pmatrix} + {\rm h.c.} \;.
	\end{eqnarray}
In the flavor basis where the charged-lepton mass matrix is diagonal, the diagonalization of the $6\times 6$ neutrino mass matrix in Eq.~(\ref{eq:mass_term}) leads to the mass spectrum of six Majorana neutrinos and the $6\times 6$ unitary flavor mixing matrix. Subsequently, one can investigate neutrino interactions and calculate radiative corrections to physical observables.
    
    \item For $\Lambda^{}_{\rm R} \gg \Lambda^{}_{\rm EW}$, two well-separated energy scales are present in the UV complete theory. In this case, it is more convenient to follow the approach of effective field theories, which have been demonstrated as a powerful tool to deal with multi-scale problems. Taking right-handed neutrinos to be heavy, i.e., $\Lambda^{}_{\rm R} = {\cal O}({\bf m}_{\rm R}^{}) \gg \Lambda^{}_{\rm EW}$, one can integrate them out in the path-integral formalism and construct a low-energy effective theory that is defined by the SM Lagrangian and a series of effective operators of mass dimensions higher than four. Up to dimension-six, the gauge-invariant Lagrangian of the seesaw effective field theory (SEFT) at the tree level is given by~\cite{Broncano:2002rw,Broncano:2003fq}
        \begin{eqnarray}\label{eq:L_SEFT}
          {\cal L}^{}_{\rm SEFT} &=& {\cal L}^{}_{\rm SM} + \frac{1}{2}\left({\bf C}^{}_5 {\cal O}^{}_5 + {\rm h.c.}\right) + {\bf C}^{}_6 {\cal O}^{}_6 \; ,
        \end{eqnarray}
        where ${\cal O}^{}_5 \equiv \overline{\ell^{}_{\rm L}} \widetilde{H} \widetilde{H}^{\rm T} \ell^{\rm c}_{\rm L}$ is the unique dimension-five Weinberg operator~\cite{Weinberg:1979sa} with the Wilson coefficient ${\bf C}^{}_5 = {\bf y}^{}_\nu {\bf m}^{-1}_{\rm R} {\bf y}^{\rm T}_\nu$, and the dimension-six operator is ${\cal O}^{}_6 \equiv \left(\overline{\ell^{}_{\rm L}} \widetilde{H}\right){\rm i}\slashed{\partial} \left(\widetilde{H}^\dagger \ell^{}_{\rm L}\right)$ with the Wilson coefficient ${\bf C}^{}_6 = {\bf y}^{}_\nu {\bf m}^{-2}_{\rm R} {\bf y}^\dagger_\nu$. The low-energy phenomenology of the type-I seesaw model can also be studied in the SEFT, which has been derived from the UV-complete theory, by considering the implications of high-dimensional operators and the renormalization-group (RG) running of relevant physical parameters~\cite{Chankowski:1993tx,Babu:1993qv,King:2000hk,Antusch:2001ck,Antusch:2002rr,Antusch:2005gp,Mei:2005qp,Ohlsson:2013xva} from the matching scale $\mu = \Lambda^{}_{\rm R}$ to the electroweak scale $\mu = \Lambda^{}_{\rm EW}$. At the one-loop level, the effective Lagrangian ${\cal L}^{}_{\rm SEFT}$ and the RG running of Wilson coefficients have been discussed in Refs.~\cite{Zhang:2021jdf,Wang:2023bdw} in a systematic way. After the spontaneous gauge symmetry breaking, the effective Majorana mass term for ordinary neutrinos arises from the Weinberg operator and the associated mass matrix is ${\bf m}^{}_\nu = - {\bf m}^{}_{\rm D} {\bf m}^{-1}_{\rm R} {\bf m}^{\rm T}_{\rm D}$, which can also be obtained from Eq.~(\ref{eq:mass_term}) via block diagonalization of the overall mass matrix in the leading-order approximation with ${\cal O}({\bf m}^{}_{\rm D} {\bf m}^{-1}_{\rm R}) \ll 1$.
\end{itemize}

In the type-I seesaw model, the smallness of ordinary neutrino masses can be ascribed to the largeness of heavy Majorana neutrino masses. Moreover, the out-of-equilibrium and CP-violating decays of heavy Majorana neutrinos in the early Universe generate lepton number asymmetries, which can be partially converted into baryon number asymmetry via the sphaleron process~\cite{Fukugita:1986hr}. These two salient features motivate us to carry out a complete one-loop renormalization of the type-I seesaw model, which is indispensable for testing it with future precision data. In addition to the existing experimental data that are well consistent with the SM predictions beyond the tree level, the next-generation neutrino oscillation experiments are about to measure the oscillation parameters at the percent or even sub-percent level~\cite{JUNO:2022mxj,Capozzi:2025wyn}. At the same time, the lepton-flavor-violating processes (e.g., $\mu \to e + \gamma$ decays and $\mu$-$e$ conversions)~\cite{Calibbi:2017uvl,Fernandez-Martinez:2024bxg}, neutrinoless double-beta decays of even-even nuclei~\cite{Rodejohann:2011mu,Dolinski:2019nrj} and future high-energy colliders~\cite{Cai:2017mow,Fernandez-Martinez:2023phj} will definitely offer more complementary data for this purpose.

In this work, starting from the classical Lagrangian in Eq.~(\ref{eq:L_SS}), we find out the neutrino mass spectrum and neutrino interactions, from which the corresponding Feynman rules are then derived. Next, the one-loop renormalization is performed in the modified minimal-subtraction ($\overline{\rm MS}$) scheme~\cite{tHooft:1973mfk,Weinberg:1973xwm,Bardeen:1978yd}.\footnote{Compared to the on-shell scheme which is also widely adopted in the renormalization of the electroweak theory, the $\overline{\rm MS}$ scheme is more practical for calculating higher-order corrections. Its advantages are particularly evident in models with different mass scales when adopting the approach of effective field theories. Meanwhile, the RG running behaviors for physical parameters should be derived in the $\overline{\rm MS}$ scheme.} The physical parameters are chosen as the fine-structure constant $\alpha(\mu)$, $W$- and $Z$-boson masses $\{m_W^{}(\mu),m_Z^{}(\mu)\}$, the Higgs-boson mass $m_h^{}(\mu)$, fermion masses $m_f^{}(\mu)$, and the flavor mixing matrices $\{{\bf V}_{}^{\rm CKM}(\mu), {\bf V}(\mu), {\bf R}(\mu)\}$ for quarks, light and heavy Majorana neutrinos, respectively, where $\mu$ denotes the 't Hooft mass scale. With those running parameters, the one-loop self-energies of charged leptons and massive Majorana neutrinos are calculated. The explicit expressions of all the counterterms for wave functions, lepton masses and the leptonic flavor mixing matrix are given as well. The full set of RG equations (RGEs) for fermion masses, flavor mixing matrices and all the original SM parameters are derived. Finally, we adopt the Euler-like parametrization of the leptonic flavor mixing matrix~\cite{Xing:2007zj,Xing:2011ur}, and derive the RGEs for the mixing angles and CP-violating phases, which can be used for precise calculations of physical observables in the type-I seesaw model.

The remaining part of this paper is organized as follows. In Sec.~\ref{sec:typeI} we examine the classical Lagrangian of the type-I seesaw model and list the Feynman rules. The basic strategy for one-loop renormalization in the $\overline{\rm MS}$ is outlined in Sec.~\ref{sec:renorm}, where the input parameters and the tadpole scheme are chosen and explained. The explicit results of the counterterms for wave functions, lepton masses and mixing matrix elements are obtained in Sec.~\ref{sec:RGE}, where the RGEs for all physical parameters are also derived. We summarize our main results in Sec.~\ref{sec:sum}. Analytical expressions for the self-energy corrections of Majorana neutrinos and charged leptons in the $R_\xi^{}$ gauge are given in Appendix~\ref{app:1-loop}.

\section{Neutrino Interactions and Feynman Rules}
\label{sec:typeI}
	
In this section, we examine the implications of the classical Lagrangian in Eq.~(\ref{eq:L_SS}) to establish our notations for later discussions. The $6\times 6$ neutrino mass matrix in Eq.~(\ref{eq:mass_term}) can be diagonalized by a $6\times6$ unitary matrix ${\cal U}$, which is further divided into four $3\times 3$ sub-matrices as 
\begin{eqnarray}
		\label{eq:VRSU}
		{\cal U} = \begin{pmatrix}
			{\bf V} & {\bf R} \\ {\bf S} & {\bf U}
		\end{pmatrix} \;,
\end{eqnarray}
where ${\bf V}$, ${\bf R}$, ${\bf S}$ and ${\bf U}$ are non-unitary but constrained by the unitarity conditions ${\bf V}^\dagger {\bf V} + {\bf S}^\dagger {\bf S} = {\bf R}^\dagger {\bf R} + {\bf U}^\dagger {\bf U} = {\bf V}{\bf V}^\dagger + {\bf R}{\bf R}^\dagger = {\bf S}{\bf S}^\dagger + {\bf U}{\bf U}^\dagger = {\bf 1}$ and ${\bf V}^\dagger {\bf R} + {\bf S}^\dagger {\bf U} = {\bf V} {\bf S}^\dagger + {\bf R}{\bf U}^\dagger = {\bf 0}$. To be explicit, we have
\begin{eqnarray}
		\label{eq:diagonal_mass_matrix}
		\begin{pmatrix} {\bf V} & {\bf R} \\ {\bf S} & {\bf U} \end{pmatrix}^\dagger
		\begin{pmatrix} {\bf 0} & {\bf m}_{\rm D}^{} \\ {\bf m}_{\rm D}^{\rm T} & {\bf m}_{\rm R}^{} \end{pmatrix}
		\begin{pmatrix} {\bf V} & {\bf R} \\ {\bf S} & {\bf U} \end{pmatrix}^* 
		= 
		\begin{pmatrix} \widehat{\bf m} & {\bf 0} \\ {\bf 0} & \widehat{\bf M} \end{pmatrix} \equiv {\cal M} \;,
	\end{eqnarray}
where $\widehat{\bf m} \equiv {\rm diag}\left\{m_1^{}, m_2^{}, m_3^{}\right\}$ and $\widehat{\bf M} \equiv {\rm diag}\left\{M_1^{}, M_2^{}, M_3^{}\right\}$ represent the masses of light and heavy Majorana neutrinos, respectively. Note that all the matrices of six rows or columns, such as ${\cal U}$ and ${\cal M}$, will be denoted by calligraphic Latin letters. From Eq.~(\ref{eq:diagonal_mass_matrix}), we can obtain the {\it exact} seesaw relation
	\begin{eqnarray}
		\label{eq:seesaw_relation}
		{\bf V} \widehat{\bf m} {\bf V}^{\rm T} + {\bf R} \widehat{\bf M} {\bf R}^{\rm T} = {\bf 0} \;,
	\end{eqnarray}
and express the Dirac neutrino mass matrix ${\bf m}^{}_{\rm D}$ or equivalently the Yukawa coupling matrix ${\bf y}^{}_\nu$ in terms of mass eigenvalues and the mixing matrices, i.e.,
	\begin{eqnarray}
		\label{eq:m_D}
		\quad {\bf m}_{\rm D}^{} = {\bf V} \widehat{\bf m} {\bf S}^{\rm T} + {\bf R} \widehat{\bf M} {\bf U}^{\rm T} \;.
	\end{eqnarray}
	One can define the corresponding mass eigenstates for light Majorana neutrinos $\widehat{\nu} = \widehat{\nu}_{\rm L}^{} + \widehat{\nu}_{\rm L}^{\rm c}$ and heavy Majorana neutrinos $\widehat{N} = \widehat{N}_{\rm R}^{\rm c} + \widehat{N}_{\rm R}^{}$. Apparently, they satisfy the Majorana conditions $\widehat{\nu}^{\rm c} = \widehat{\nu}$ and $\widehat{N}^{\rm c} = \widehat{N}$. Notice that the neutrino flavor and mass eigenfields are connected through the following relations
	\begin{eqnarray}
		\label{eq:flavor_mass}
		\begin{pmatrix} \nu_{\rm L}^{} \\ N_{\rm R}^{\rm c} \end{pmatrix} 
		= 
		\begin{pmatrix} {\bf V} & {\bf R} \\ {\bf S} & {\bf U} \end{pmatrix}
		\begin{pmatrix} \widehat{\nu}_{\rm L}^{} \\ \widehat{N}_{\rm R}^{\rm c} \end{pmatrix} \;, \quad
		\begin{pmatrix} \nu_{\rm L}^{\rm c} \\ N_{\rm R}^{} \end{pmatrix} 
		= 
		\begin{pmatrix}	{\bf V} & {\bf R} \\ {\bf S} & {\bf U} \end{pmatrix}^* 
		\begin{pmatrix} \widehat{\nu}_{\rm L}^{\rm c} \\ \widehat{N}_{\rm R}^{} \end{pmatrix} \;.
	\end{eqnarray}
For later convenience, we label the light neutrino mass eigenfields as $\chi_i^{} \equiv \widehat{\nu}_i^{}$ (for $i=1,2,3$) and heavy ones as $\chi_{i+3}^{} \equiv \widehat{N}_i^{}$ (for $i=1,2,3$). Meanwhile, the corresponding neutrino masses can be labeled as $\widehat{m}_i^{} = m_i^{}$ (for $i=1,2,3$) for light ones, and $\widehat{m}_{i+3}^{} = M_{i}^{}$ (for $i=1,2,3$) for heavy ones.
	
With the help of the relations in Eqs.~(\ref{eq:seesaw_relation})-(\ref{eq:m_D}) and the unitarity conditions of ${\cal U}$, we can explicitly write down the interactions of massive Majorana neutrinos. In the flavor basis,
\begin{eqnarray}
{\cal L}_{\rm CC}^{} &=& \frac{g}{\sqrt{2}} \overline{l_{\rm L}^{}} \gamma^\mu P_{\rm L}^{} \nu_{\rm L}^{} W_\mu^- + {\rm h.c.} \;, \nonumber \\
{\cal L}_{\rm NC}^{} &=& \frac{g}{2c} \overline{\nu_{\rm L}^{}} \gamma^\mu P_{\rm L}^{} \nu_{\rm L}^{} Z_\mu  \;, \nonumber \\
{\cal L}_{h}^{} &=& - \frac{1}{\sqrt{2}} \overline{\nu_{\rm L}^{}} {\bf y}_\nu^{} N_{\rm R}^{} h + {\rm h.c.} \;, \nonumber \\
{\cal L}_{\phi^0}^{} &=& \frac{{\rm i}}{\sqrt{2}} \overline{\nu_{\rm L}^{}} {\bf y}_\nu^{} N_{\rm R}^{} \phi^0 + {\rm h.c.} \;, \nonumber \\
{\cal L}_{\rm \phi^\pm}^{} &=& \overline{l_{\rm L}^{}} {\bf y}_\nu^{} N_{\rm R}^{} \phi^- - \overline{\nu_{\rm L}^{}} {\bf y}_\alpha^{} l_{\rm R}^{} \phi^+ + {\rm h.c.} \;, 
\end{eqnarray}
refer to the ordinary charged-current (CC) and neutral-current (NC) interactions with weak gauge bosons in the leptonic sector, and neutrino interactions with the Higgs boson $h$, the Goldstone bosons $\phi^0$ and $\phi^\pm$, where $g$ is the ${\rm SU}(2)_{\rm L}^{}$ gauge coupling and $c\equiv \cos\theta_{\rm w}^{}$ with $\theta_{\rm w}^{}$ being the weak mixing angle, $P_{\rm L,R}^{} \equiv \left(1\mp \gamma^5\right)/2$ are the left- and right-handed chiral projection operators. In the mass basis, those interaction terms can be recast into~\cite{Pilaftsis:1991ug}
	\begin{eqnarray}
		{\cal L}_{\rm CC}^{} &=& \frac{g}{\sqrt{2}} \overline{l_\alpha^{}} \gamma^\mu P_{\rm L}^{} {\cal B}_{\alpha i}^{} \chi_i^{} W_\mu^- + {\rm h.c.} \;, \nonumber \\
		{\cal L}_{\rm NC}^{} &=& \frac{g}{4c} \overline{\chi_i^{}} \gamma^\mu \left(P_{\rm L}^{} {\cal C}_{ij}^{} - P_{\rm R}^{} {\cal C}_{ij}^* \right) \chi_j^{} Z_\mu \;, \nonumber \\
		{\cal L}_h^{} &=& - \frac{g}{4 m_W^{}} h \overline{\chi_i^{}} \left[ \left(\widehat{m}_j^{} P_{\rm R}^{} + \widehat{m}_i^{} P_{\rm L}^{}\right) {\cal C}_{ij}^{} + \left(\widehat{m}_i^{} P_{\rm R}^{} + \widehat{m}_j^{} P_{\rm L}^{}  \right) {\cal C}^*_{ij} \right] \chi_j^{} \;, \nonumber \\
		{\cal L}_{\phi^\pm}^{} &=& \frac{g}{\sqrt{2} m_W^{}} \overline{l_\alpha^{}} \left(\widehat{m}_i^{} P_{\rm R}^{} - m_\alpha^{}  P_{\rm L}^{}  \right) {\cal B}_{\alpha i}^{} \chi_i^{} \phi^- + {\rm h.c.} \;, \nonumber \\
		{\cal L}_{\phi^0}^{} &=& \frac{{\rm i} g}{4 m_W^{}} \phi^0 \overline{\chi_i^{}} \left[\left(\widehat{m}_j^{} P_{\rm R}^{} - \widehat{m}_i^{} P_{\rm L}^{}\right) {\cal C}_{ij}^{} +  \left(\widehat{m}_i^{} P_{\rm R}^{} - \widehat{m}_j^{} P_{\rm L}^{}\right) {\cal C}^*_{ij} \right] \chi_j^{} \;,
	\end{eqnarray}
where $\alpha = e,\mu,\tau$ denote three charged-lepton flavors and $i,j=1,2,\cdots,6$ run over six Majorana neutrino mass eigenstates. The $W$-boson mass is $m_W^{} = g v^{}_0/2$, and the Yukawa coupling matrix for charged leptons is ${\bf y}_\alpha^{} = g {\bf m}_{\alpha}^{} / (\sqrt{2} m_W^{})$ with ${\bf m}_\alpha^{} = {\rm diag} \left\{m_e^{}, m_\mu^{}, m_\tau^{}\right\}$ being the diagonal mass matrix. Moreover, the $3 \times 6$ flavor mixing matrix ${\cal B}$ and the $6\times 6$ Hermitian matrix ${\cal C}$ have been introduced, i.e.,
	\begin{eqnarray}
		{\cal B} \equiv \begin{pmatrix} {\bf V} & {\bf R} \end{pmatrix} \;, \quad {\cal C} \equiv {\cal B}^\dagger {\cal B} = \begin{pmatrix} {\bf V}^\dagger {\bf V} & {\bf V}^\dagger {\bf R} \\ {\bf R}^\dagger {\bf V} & {\bf R}^\dagger {\bf R} \end{pmatrix} \;,
	\end{eqnarray}
where ${\cal B} {\cal B}^\dagger = {\bf V} {\bf V}^\dagger + {\bf R} {\bf R}^\dagger = {\bf 1}$ holds according to the unitarity condition. Meanwhile, the seesaw relation in Eq.~(\ref{eq:seesaw_relation}) can be further expressed as ${\cal B} {\cal M} {\cal B}^{\rm T} = {\bf 0}$. Based on the definitions and the observations above, one can immediately verify the following identities
\begin{eqnarray}
\sum_{k=1}^6 {\cal B}_{\alpha k}^{} {\cal B}_{\beta k}^* = \delta_{\alpha\beta}^{} \;, \quad
\sum_{k=1}^6 {\cal B}_{\alpha k}^{} {\cal C}_{k i}^{} = {\cal B}_{\alpha i}^{} \;, \quad
\sum_{k=1}^6 {\cal C}_{ik}^{} {\cal C}_{kj}^{} = {\cal C}_{ij}^{} \;, 
\end{eqnarray}
among the matrix elements of ${\cal B}$ and ${\cal C}$; and
\begin{eqnarray}
\sum_{k=1}^6 \widehat{m}_k^{} {\cal B}_{\alpha k}^{} {\cal B}_{\beta k}^{} = 0 \;, \quad
\sum_{k=1}^6 \widehat{m}_k^{} {\cal B}_{\alpha k}^{} {\cal C}_{ki}^{*} = 0 \;, \quad 
\sum_{k=1}^6 \widehat{m}_k^{} {\cal C}_{ik}^{} {\cal C}_{kj}^* = 0 \;,
\end{eqnarray}
where neutrino masses are also involved.

Except for the neutrino sector, all other terms in the Lagrangian of the type-I seesaw model are the same as those in the SM, for which the Feynman rules in the $R_\xi^{}$ gauge can be found in Ref.~\cite{Denner:2019vbn}. We focus on the interactions including Majorana neutrinos in the subsequent discussions. The general procedure to derive the Feynman rules for Majorana particles in the theory can be found in Ref.~\cite{Denner:1992vza}. First, the propagator of massive Majorana neutrinos is
	\begin{eqnarray}
		\begin{tikzpicture}[baseline=0]
			\begin{feynman}
				% 定义顶点
				\vertex (a) {$\chi_i^{}$};
				\vertex [right=3cm of a] (b) {$\chi_i^{}$};
				
				% 绘制费曼图并标注动量
				\diagram* {
					(a) -- [momentum= $p_i^{}$] (b) ,
				};
			\end{feynman}
		\end{tikzpicture} = \frac{{\rm i} (\slashed{p}_i^{} + \widehat{m}_i^{})}{p_i^2 - \widehat{m}_i^2} \;.
	\end{eqnarray}
For Dirac fermions, each fermion line carries an arrow describing the {\it fermion number flow} to display the flow of charge, namely, to distinguish between particles and antiparticles. For Majorana fermions, since particles and antiparticles are indistinguishable, no such arrows are needed, but the arrow of {\it fermion flow} will be drawn in gray along the fermion line attached to each vertex. For the CC interaction, the corresponding Feynman rules with various directions of the fermion flow and fermion number flow are collected below:
	\begin{eqnarray}
		&\begin{tikzpicture}[baseline=0]
			\begin{feynman}
				% 顶点定义
				\vertex (a) {$\chi_i^{}$};
				\vertex [right = 0.7 cm of a] (b);
				\vertex [right = 0.7 cm of b] (c);
				\vertex [above right= 0.6 cm of c] (h);
				\vertex [above right= 0.6 cm of h] (d) {$l_\alpha^{-}$};
				\vertex [below right =1.0 cm of c] (e) {$W_\mu$};
				\vertex [above = 0.2 cm of b] (f);
				\vertex [above left = 0.2 cm of h] (g);
				
				% 基础费曼图
				\diagram* {
					(a) --  (c) -- [fermion] (d),
					(c) -- [photon] (e),    };
				
				% 额外弧形费米流（Majorana自能修正）
				\draw [fermion, gray, half right] 
				(f) to [out=-20, in=-160, looseness = 1] (g);
			\end{feynman}
		\end{tikzpicture} \displaystyle = \frac{{\rm i} g}{\sqrt{2}} \gamma_\mu P_{\rm L}^{} {\cal B}_{\alpha i}^{} \;, 
		\qquad
		\begin{tikzpicture}[baseline=0]
			\begin{feynman}
				% 顶点定义
				\vertex (a) {$\chi_i^{}$};
				\vertex [right = 0.7 cm of a] (b);
				\vertex [right = 0.7 cm of b] (c);
				\vertex [above right= 0.6 cm of c] (h);
				\vertex [above right= 0.6 cm of h] (d) {$l_\alpha^{+}$};
				\vertex [below right =1.0 cm of c] (e) {$W_\mu$};
				\vertex [above = 0.2 cm of b] (f);
				\vertex [above left = 0.2 cm of h] (g);
				
				% 基础费曼图
				\diagram* {
					(a) --  (c) -- [anti fermion] (d),
					(c) -- [photon] (e),    };
				
				% 额外弧形费米流（Majorana自能修正）
				\draw [anti fermion, gray, half right] 
				(f) to [out=-20, in=-160, looseness = 1] (g);
			\end{feynman}
		\end{tikzpicture} = \frac{{\rm i} g}{\sqrt{2}} \gamma_\mu P_{\rm L}^{} {\cal B}_{\alpha i}^{*} \;, \nonumber \\
		&\begin{tikzpicture}[baseline=0]
			\begin{feynman}
				% 顶点定义
				\vertex (a) {$\chi_i^{}$};
				\vertex [right = 0.7 cm of a] (b);
				\vertex [right = 0.7 cm of b] (c);
				\vertex [above right= 0.6 cm of c] (h);
				\vertex [above right= 0.6 cm of h] (d) {$l_\alpha^{-}$};
				\vertex [below right =1.0 cm of c] (e) {$W_\mu$};
				\vertex [above = 0.2 cm of b] (f);
				\vertex [above left = 0.2 cm of h] (g);
				
				% 基础费曼图
				\diagram* {
					(a) --  (c) -- [fermion] (d),
					(c) -- [photon] (e),    };
				
				% 额外弧形费米流（Majorana自能修正）
				\draw [anti fermion, gray, half right] 
				(f) to [out=-20, in=-160, looseness = 1] (g);
			\end{feynman}
		\end{tikzpicture} \displaystyle = - \frac{{\rm i} g}{\sqrt{2}} \gamma_\mu P_{\rm R}^{} {\cal B}_{\alpha i}^{} \;,
		\qquad
		\begin{tikzpicture}[baseline=0]
			\begin{feynman}
				% 顶点定义
				\vertex (a) {$\chi_i^{}$};
				\vertex [right = 0.7 cm of a] (b);
				\vertex [right = 0.7 cm of b] (c);
				\vertex [above right= 0.6 cm of c] (h);
				\vertex [above right= 0.6 cm of h] (d) {$l_\alpha^{+}$};
				\vertex [below right =1.0 cm of c] (e) {$W_\mu$};
				\vertex [above = 0.2 cm of b] (f);
				\vertex [above left = 0.2 cm of h] (g);
				
				% 基础费曼图
				\diagram* {
					(a) --  (c) -- [anti fermion] (d),
					(c) -- [photon] (e),    };
				
				% 额外弧形费米流（Majorana自能修正）
				\draw [fermion, gray, half right] 
				(f) to [out=-20, in=-160, looseness = 1] (g);
			\end{feynman}
		\end{tikzpicture} = - \frac{{\rm i} g}{\sqrt{2}} \gamma_\mu P_{\rm R}^{} {\cal B}_{\alpha i}^{*} \;. 
	\end{eqnarray}
The Feynman rules for neutrino interactions with the charged Goldstone bosons $\phi^\pm$ are
	\begin{eqnarray}
		\begin{array}{ll}
		 &\begin{tikzpicture}[baseline=0]
			\begin{feynman}
				% 顶点定义
				\vertex (a) {$\chi_i^{}$};
				\vertex [right = 0.7 cm of a] (b);
				\vertex [right = 0.7 cm of b] (c);
				\vertex [above right= 0.6 cm of c] (h);
				\vertex [above right= 0.6 cm of h] (d) {$l_\alpha^{-}$};
				\vertex [below right =1.0 cm of c] (e) {$\phi^+$};
				\vertex [above = 0.2 cm of b] (f);
				\vertex [above left = 0.2 cm of h] (g);
				
				% 基础费曼图
				\diagram* {
					(a) --  (c) -- [fermion] (d),
					(c) -- [scalar] (e),    };
				
				% 额外弧形费米流（Majorana自能修正）
				\draw [fermion, gray, half right] 
				(f) to [out=-20, in=-160, looseness = 1] (g);
			\end{feynman}
		\end{tikzpicture} \\
		 = & \displaystyle \frac{{\rm i} g}{\sqrt{2} m_W^{}} \left(\widehat{m}_i^{} P_{\rm R}^{} - m_\alpha^{} P_{\rm L}^{}\right) {\cal B}_{\alpha i}^{} \;, \end{array} \qquad
		\begin{array}{ll}
		&\begin{tikzpicture}[baseline=0]
			\begin{feynman}
				% 顶点定义
				\vertex (a) {$\chi_i^{}$};
				\vertex [right = 0.7 cm of a] (b);
				\vertex [right = 0.7 cm of b] (c);
				\vertex [above right= 0.6 cm of c] (h);
				\vertex [above right= 0.6 cm of h] (d) {$l_\alpha^{+}$};
				\vertex [below right =1.0 cm of c] (e) {$\phi^-$};
				\vertex [above = 0.2 cm of b] (f);
				\vertex [above left = 0.2 cm of h] (g);
				
				% 基础费曼图
				\diagram* {
					(a) --  (c) -- [anti fermion] (d),
					(c) -- [scalar] (e),    };
				
				% 额外弧形费米流（Majorana自能修正）
				\draw [anti fermion, gray, half right] 
				(f) to [out=-20, in=-160, looseness = 1] (g);
			\end{feynman}
		\end{tikzpicture} \\
		= & \displaystyle \frac{{\rm i} g}{\sqrt{2} m_W^{}} \left(\widehat{m}_i^{} P_{\rm L}^{} - m_\alpha^{} P_{\rm R}^{}\right) {\cal B}_{\alpha i}^{*} \;, \end{array} \nonumber \\
		\begin{array}{ll}
		& \begin{tikzpicture}[baseline=0]
			\begin{feynman}
				% 顶点定义
				\vertex (a) {$\chi_i^{}$};
				\vertex [right = 0.7 cm of a] (b);
				\vertex [right = 0.7 cm of b] (c);
				\vertex [above right= 0.6 cm of c] (h);
				\vertex [above right= 0.6 cm of h] (d) {$l_\alpha^{-}$};
				\vertex [below right =1.0 cm of c] (e) {$\phi^+$};
				\vertex [above = 0.2 cm of b] (f);
				\vertex [above left = 0.2 cm of h] (g);
				
				% 基础费曼图
				\diagram* {
					(a) --  (c) -- [fermion] (d),
					(c) -- [scalar] (e),    };
				
				% 额外弧形费米流（Majorana自能修正）
				\draw [anti fermion, gray, half right] 
				(f) to [out=-20, in=-160, looseness = 1] (g);
			\end{feynman}
		\end{tikzpicture} \\		
		= & \displaystyle \frac{{\rm i} g}{\sqrt{2} m_W^{}} \left(\widehat{m}_i^{} P_{\rm R}^{} - m_\alpha^{} P_{\rm L}^{}\right) {\cal B}_{\alpha i}^{} \;, \end{array}
		\qquad
		\begin{array}{ll}
		& \begin{tikzpicture}[baseline=0]
			\begin{feynman}
				% 顶点定义
				\vertex (a) {$\chi_i^{}$};
				\vertex [right = 0.7 cm of a] (b);
				\vertex [right = 0.7 cm of b] (c);
				\vertex [above right= 0.6 cm of c] (h);
				\vertex [above right= 0.6 cm of h] (d) {$l_\alpha^{+}$};
				\vertex [below right =1.0 cm of c] (e) {$\phi^-$};
				\vertex [above = 0.2 cm of b] (f);
				\vertex [above left = 0.2 cm of h] (g);
				
				% 基础费曼图
				\diagram* {
					(a) --  (c) -- [anti fermion] (d),
					(c) -- [scalar] (e),    };
				
				% 额外弧形费米流（Majorana自能修正）
				\draw [fermion, gray, half right] 
				(f) to [out=-20, in=-160, looseness = 1] (g);
			\end{feynman}
		\end{tikzpicture} \\
		= & \displaystyle \frac{{\rm i} g}{\sqrt{2} m_W^{}} \left(\widehat{m}_i^{} P_{\rm L}^{} - m_\alpha^{} P_{\rm R}^{}\right) {\cal B}_{\alpha i}^{*} \;. \end{array}
	\end{eqnarray}
	
For neutrino interactions with neutral gauge or scalar bosons, the Feynman rules can be derived in the same way. However, an extra factor of two should be taken into consideration, as two types of contractions of Majorana neutrino fields yield the same result. For the NC interaction, the Feynman rules are
	\begin{eqnarray}
		\begin{array}{ll}
			&\begin{tikzpicture}[baseline=0]
				\begin{feynman}
					% 顶点定义
					\vertex (a) {$\chi_j^{}$};
					\vertex [right = 0.7 cm of a] (b);
					\vertex [right = 0.7 cm of b] (c);
					\vertex [above right= 0.5 cm of c] (h);
					\vertex [above right= 0.5 cm of h] (d) {$\chi_i^{}$};
					\vertex [below right =1.0 cm of c] (e) {$Z_\mu$};
					\vertex [above = 0.2 cm of b] (f);
					\vertex [above left = 0.2 cm of h] (g);
					
					% 基础费曼图
					\diagram* {
						(a) --  (c) --  (d),
						(c) -- [boson] (e),    };
					
					% 额外弧形费米流（Majorana自能修正）
					\draw [fermion, gray, half right] 
					(f) to [out=-20, in=-160, looseness = 1] (g);
				\end{feynman}
			\end{tikzpicture} \\
			= & \displaystyle \frac{{\rm i} g}{2c} \gamma_\mu \left(P_{\rm L}^{} {\cal C}_{ij}^{} - P_{\rm R}^{} {\cal C}_{ij}^*\right) \;, \end{array} \qquad
		\begin{array}{ll}
			&\begin{tikzpicture}[baseline=0]
				\begin{feynman}
					% 顶点定义
					\vertex (a) {$\chi_j^{}$};
					\vertex [right = 0.7 cm of a] (b);
					\vertex [right = 0.7 cm of b] (c);
					\vertex [above right= 0.5 cm of c] (h);
					\vertex [above right= 0.5 cm of h] (d) {$\chi_i^{}$};
					\vertex [below right =1.0 cm of c] (e) {$Z_\mu$};
					\vertex [above = 0.2 cm of b] (f);
					\vertex [above left = 0.2 cm of h] (g);
					
					% 基础费曼图
					\diagram* {
						(a) --  (c) --  (d),
						(c) -- [boson] (e),    };
					
					% 额外弧形费米流（Majorana自能修正）
					\draw [anti fermion, gray, half right] 
					(f) to [out=-20, in=-160, looseness = 1] (g);
				\end{feynman}
			\end{tikzpicture} \\
			= & \displaystyle \frac{{\rm i} g}{2c} \gamma_\mu \left(P_{\rm L}^{} {\cal C}_{ji}^{} - P_{\rm R}^{} {\cal C}_{ji}^*\right) \;. \end{array}
	\end{eqnarray}
One may notice that the Feynman rule of the latter vertex can be simply obtained from that of the former by exchanging $i\leftrightarrow j$ due to the symmetric interaction Lagrangian. The interaction with the Higgs boson is described by the following vertices
	\begin{eqnarray}
		\begin{tikzpicture}[baseline=0]
			\begin{feynman}
				% 顶点定义
				\vertex (a) {$\chi_j^{}$};
				\vertex [right = 0.7 cm of a] (b);
				\vertex [right = 0.7 cm of b] (c);
				\vertex [above right= 0.5 cm of c] (h);
				\vertex [above right= 0.5 cm of h] (d) {$\chi_i^{}$};
				\vertex [below right =1.0 cm of c] (e) {$h$};
				\vertex [above = 0.2 cm of b] (f);
				\vertex [above left = 0.2 cm of h] (g);
				
				% 基础费曼图
				\diagram* {
					(a) --  (c) --  (d),
					(c) -- [scalar] (e),    };
				
				% 额外弧形费米流（Majorana自能修正）
				\draw [fermion, gray, half right] 
				(f) to [out=-20, in=-160, looseness = 1] (g);
			\end{feynman}
		\end{tikzpicture}
		&=& - \frac{{\rm i} g}{2 m_W^{}} \left[\left(\widehat{m}_j^{} P_{\rm R}^{} + \widehat{m}_i^{} P_{\rm L}^{}\right) {\cal C}_{ij}^{} + \left(\widehat{m}_i^{} P_{\rm R}^{} + \widehat{m}_j^{} P_{\rm L}^{}\right) {\cal C}_{ij}^*\right] \;, \nonumber \\
		\begin{tikzpicture}[baseline=0]
			\begin{feynman}
				% 顶点定义
				\vertex (a) {$\chi_j^{}$};
				\vertex [right = 0.7 cm of a] (b);
				\vertex [right = 0.7 cm of b] (c);
				\vertex [above right= 0.5 cm of c] (h);
				\vertex [above right= 0.5 cm of h] (d) {$\chi_i^{}$};
				\vertex [below right =1.0 cm of c] (e) {$h$};
				\vertex [above = 0.2 cm of b] (f);
				\vertex [above left = 0.2 cm of h] (g);
				
				% 基础费曼图
				\diagram* {
					(a) --  (c) --  (d),
					(c) -- [scalar] (e),    };
				
				% 额外弧形费米流（Majorana自能修正）
				\draw [anti fermion, gray, half right] 
				(f) to [out=-20, in=-160, looseness = 1] (g);
			\end{feynman}
		\end{tikzpicture}
		&=& - \frac{{\rm i} g}{2 m_W^{}} \left[\left(\widehat{m}_i^{} P_{\rm R}^{} + \widehat{m}_j^{} P_{\rm L}^{}\right) {\cal C}_{ji}^{} + \left(\widehat{m}_j^{} P_{\rm R}^{} + \widehat{m}_i^{} P_{\rm L}^{}\right) {\cal C}_{ji}^*\right] \;.
	\end{eqnarray}
Finally, the vertices involving the neutral Goldstone boson $\phi^0$ read
	\begin{eqnarray}
		\begin{tikzpicture}[baseline=0]
			\begin{feynman}
				% 顶点定义
				\vertex (a) {$\chi_j^{}$};
				\vertex [right = 0.7 cm of a] (b);
				\vertex [right = 0.7 cm of b] (c);
				\vertex [above right= 0.5 cm of c] (h);
				\vertex [above right= 0.5 cm of h] (d) {$\chi_i^{}$};
				\vertex [below right =1.0 cm of c] (e) {$\phi^0$};
				\vertex [above = 0.2 cm of b] (f);
				\vertex [above left = 0.2 cm of h] (g);
				
				% 基础费曼图
				\diagram* {
					(a) --  (c) --  (d),
					(c) -- [scalar] (e),    };
				
				% 额外弧形费米流（Majorana自能修正）
				\draw [fermion, gray, half right] 
				(f) to [out=-20, in=-160, looseness = 1] (g);
			\end{feynman}
		\end{tikzpicture}
		&=& - \frac{g}{2 m_W^{}} \left[\left(\widehat{m}_j^{} P_{\rm R}^{} - \widehat{m}_i^{} P_{\rm L}^{}\right) {\cal C}_{ij}^{} + \left(\widehat{m}_i^{} P_{\rm R}^{} - \widehat{m}_j^{} P_{\rm L}^{}\right) {\cal C}_{ij}^*\right] \;, \nonumber \\
		\begin{tikzpicture}[baseline=0]
			\begin{feynman}
				% 顶点定义
				\vertex (a) {$\chi_j^{}$};
				\vertex [right = 0.7 cm of a] (b);
				\vertex [right = 0.7 cm of b] (c);
				\vertex [above right= 0.5 cm of c] (h);
				\vertex [above right= 0.5 cm of h] (d) {$\chi_i^{}$};
				\vertex [below right =1.0 cm of c] (e) {$\phi^0$};
				\vertex [above = 0.2 cm of b] (f);
				\vertex [above left = 0.2 cm of h] (g);
				
				% 基础费曼图
				\diagram* {
					(a) --  (c) --  (d),
					(c) -- [scalar] (e),    };
				
				% 额外弧形费米流（Majorana自能修正）
				\draw [anti fermion, gray, half right] 
				(f) to [out=-20, in=-160, looseness = 1] (g);
			\end{feynman}
		\end{tikzpicture}
		&=& - \frac{g}{2 m_W^{}} \left[\left(\widehat{m}_i^{} P_{\rm R}^{} - \widehat{m}_j^{} P_{\rm L}^{}\right) {\cal C}_{ji}^{} + \left(\widehat{m}_j^{} P_{\rm R}^{} - \widehat{m}_i^{} P_{\rm L}^{}\right) {\cal C}_{ji}^*\right] \;.
	\end{eqnarray}
	These Feynman rules will be useful when calculating the one-loop self-energy corrections of charged leptons and Majorana neutrinos. It is worthwhile to emphasize that only the $3\times 3$ matrices ${\bf V}$ and ${\bf R}$, or equivalently ${\cal B}$, are relevant for neutrino interactions, implying that other two matrices ${\bf S}$ and ${\bf U}$ in the mixing matrix ${\cal U}$ contain no independent physical parameters. 
	
\section{Renormalization in the $\overline{\bf MS}$ scheme}
\label{sec:renorm}
	
Although the one-loop renormalization of the type-I seesaw model is similar to that of the SM~\cite{Aoki:1982ed,Bohm:1986rj,Hollik:1988ii,Denner:1991kt,Bohm:2001yx,Sirlin:2012mh,Denner:2019vbn,Huang:2023nqf}, it is worth making a few comments on the model parameters and the renormalization scheme that are different from those in the literature.
\begin{itemize}
\item First, the quantization by introducing the gauge-fixing terms and the Faddeev-Popov ghosts will not be altered at all, as right-handed neutrinos are gauge singlets in the type-I seesaw model. The original parameters appearing in the Lagrangian are: the ${\rm SU}(2)_{\rm L}^{}$ gauge coupling $g$, the ${\rm U}(1)_{\rm Y}^{}$ gauge coupling $g'$, the mass parameter $m^2$ and the Higgs quartic coupling $\lambda$ in the scalar potential $V(H) = - m^2 \left(H^\dagger H\right) + \lambda/4 \left(H^\dagger H\right)^2$, the Yukawa coupling matrices of fermions ${\bf y}_f^{}$ (where $f=u,d,\alpha,\nu$ refer to up-type quarks, down-type quarks, charged leptons and neutrinos, respectively) and the Majorana neutrino mass matrix ${\bf m}_{\rm R}^{}$. As usual, we decompose the bare parameters $\left\{g_0^{},g'_0,m_0^2,\lambda_0^{},{\bf y}_{f,0}^{},{\bf m}_{\rm R,0}^{}\right\}$ into renormalized ones $\left\{g(\mu),g'(\mu),m^2(\mu),\lambda(\mu),{\bf y}_f^{}(\mu),{\bf m}_{\rm R}^{}(\mu)\right\}$ and the corresponding counterterms. Note that we explicitly indicate the energy scale $\mu$ to emphasize the scale-dependence of the renormalized parameters in the $\overline{\rm MS}$ scheme.
		
\item In the renormalized Lagrangian, after the spontaneous gauge symmetry breaking, the weak gauge bosons and the Higgs boson become massive with masses $\left\{m_W^{}(\mu),m_Z^{}(\mu),m_h^{}(\mu)\right\}$. In the leptonic sector, as mentioned before, the physical masses of Majorana neutrinos $\widehat{m}_i^{}(\mu)$ (for $i=1,\cdots,6$) are obtained by diagonalizing the $6 \times 6$ mass matrix in Eq.(\ref{eq:mass_term}) with the unitary matrix ${\cal U}(\mu)$, while the charged-lepton masses $m_\alpha^{}(\mu)$ (for $\alpha=e,\mu,\tau$) come from the diagonal Yukawa matrix ${\bf y}_\alpha^{}(\mu)$. For quarks, we diagonalize the mass matrices ${\bf m}_{u,d}^{}(\mu)$ from ${\bf y}_{u,d}^{}(\mu)$ and obtain their masses $m_q^{}(\mu)$ (for $q=u,c,t,d,s,b$). The quark flavor mixing is described by the Cabibbo-Kobayashi-Maskawa (CKM) matrix ${\bf V}^{\rm CKM}_{}(\mu)$~\cite{Cabibbo:1963yz,Kobayashi:1973fv}. Then, the renormalized Lagrangian of the type-I seesaw model can be expressed in terms of the physical masses of gauge bosons, scalar bosons and fermions and fermion flavor mixing matrices.
		
	As in the SM, the tadpole diagrams will arise when considering higher-order corrections, and the corresponding counterterms should be introduced to get rid of the UV divergences. Hence a practical scheme for the tadpole renormalization should be specified to determine the tadpole contributions to the final results. In this work, we adopt the Fleischer-Jegerlehner tadpole scheme~\cite{Fleischer:1980ub}, in which the effective Higgs potential is expanded about the tree-level minimum $v_0^{} \equiv 2\sqrt{m_0^2/\lambda_0^{}}$, instead of the true minimum after including radiative corrections.\footnote{See, e.g., Ref.~\cite{Denner:2019vbn}, for a comprehensive review on the tadpole renormalization, where a comparative analysis of different tadpole schemes can be found.} In this tadpole scheme, for a constant $v^{}_0$, the renormalized masses $\left\{m_W^{}(\mu),m_Z^{}(\mu),m_h^{}(\mu)\right\}$ and the Dirac fermion mass matrices ${\bf m}_f^{}(\mu)$ for $f=u,d,\alpha,\nu$ are actually determined by the renormalized parameters in the generic Lagrangian via
		\begin{eqnarray}
			\label{eq:v0}
			m_W^{}(\mu) &\equiv& \frac{1}{2} g(\mu) v_0^{} \;, \nonumber \\
			m_Z^{}(\mu) &\equiv& \frac{1}{2} \sqrt{g^2(\mu) + g^{\prime 2}(\mu)} v_0^{} \;, \nonumber \\
			m_h^2(\mu) &\equiv& -m^2(\mu) + \frac{3}{4} \lambda(\mu) v_0^2 \;, \nonumber \\
			{\bf m}_f^{}(\mu) &\equiv& \frac{1}{\sqrt{2}} {\bf y}_f^{}(\mu) v_0^{} \;. 
		\end{eqnarray}
		As a result, one should include the tadpole contributions to the self-energy corrections such that both the renormalized masses and counterterms become gauge-independent. 
		
		\item Once the relations in Eq.~(\ref{eq:v0}) are implemented, the whole theory can be described by these physical parameters instead of the original parameters in the Lagrangian. To be more explicit, we choose $e(\mu)$ or equivalently the fine-structure constant $\alpha(\mu)=e^2(\mu)/(4\pi)$, the physical masses $\left\{m_W^{}(\mu),m_Z^{}(\mu),m_h^{}(\mu),m_f^{}(\mu)\right\}$ of weak gauge bosons, the Higgs boson and massive fermions, and the fermion flavor mixing matrices $\left\{{\bf V}^{\rm CKM}_{}(\mu), {\cal U}(\mu)\right\}$ as input parameters. Accordingly the counterterms are defined as 
		\begin{eqnarray}
			\label{eq:ParaCT}
			e_0^{} &=& \left(1 + \delta Z_e^{}\right) e(\mu)\;, \nonumber\\
			m_{W,0}^2 &=& m_W^2(\mu) + \delta m_W^2 \;, \nonumber\\
			m_{Z,0}^2 &=& m_Z^2(\mu) + \delta m_Z^2 \;, \nonumber\\
			m_{h,0}^2 &=& m_h^2(\mu) + \delta m_h^2 \;, \nonumber\\
			m_{f,0}^{} &=& m_f^{}(\mu) + \delta m_f^{} \;, \nonumber \\
			{\bf V}^{\rm CKM}_{0} &=& {\bf V}^{\rm CKM}_{0}(\mu) + \delta {\bf V}^{\rm CKM}_{} \;, \nonumber \\
			{\cal U}_0^{} &=& {\cal U}(\mu) + \delta {\cal U} \;,
		\end{eqnarray}
		which can be translated into the counterterms of original parameters via the relations in Eq.~(\ref{eq:v0}). Furthermore, the weak mixing angle $\theta_{\rm w}^{}(\mu)$ in the $\overline{\rm MS}$ scheme is defined via $c (\mu) \equiv \cos\theta_{\rm w}^{}(\mu) =  m_W^{}(\mu)/ m_Z^{}(\mu)$, and the weak gauge coupling can also be expressed as $g^2(\mu) = 4\pi\alpha(\mu) / s^2(\mu)$ with $s^2(\mu) \equiv \sqrt{1-c^2(\mu)} = \sqrt{1-m_W^{2}(\mu)/ m_Z^{2}(\mu)}$. 
		
Finally, we also need the wave-function counterterms to define the renormalized physical fields in the mass basis:
		\begin{eqnarray}
			\label{eq:WavFuncCT}
			W_{0\mu}^{\pm} &=& \left(1 + \frac{1}{2} \delta Z_W^{}\right) W_\mu^\pm \;, \nonumber \\
			\begin{pmatrix} Z_{0\mu}^{} \\ A_{0\mu}^{}\end{pmatrix} &=& \begin{pmatrix} 
				1 + \displaystyle \frac{1}{2} \delta Z_{ZZ}^{} & \displaystyle \frac{1}{2} \delta Z_{ZA}^{} \\ \displaystyle \frac{1}{2} \delta Z_{AZ}^{} & 1 + \displaystyle \frac{1}{2} \delta Z_{AA}^{}
			\end{pmatrix}
			\begin{pmatrix} Z_{\mu}^{} \\ A_{\mu}^{} \end{pmatrix}\;, \nonumber \\
			h_0^{} &=& \left(1 + \frac{1}{2} \delta Z_h^{}\right) h \;, \nonumber \\
			f_{i,0}^{\rm L} &=& \left(\delta_{ij}^{} + \frac{1}{2} \delta Z_{ij}^{f,{\rm L}}\right) f_j^{\rm L} \;, \nonumber \\
			f_{i,0}^{\rm R} &=& \left(\delta_{ij}^{} + \frac{1}{2} \delta Z_{ij}^{f,{\rm R}}\right) f_j^{\rm R} \;,
		\end{eqnarray}
where the subscripts ``$i$" and ``$j$" of fermion fields refer to different generations. It is worth noting that the renormalization of unphysical fields is not necessary, since they do not affect the Green's functions of physical particles and the amplitudes at the one-loop level.
		
		\item With the renormalized parameters and fields, one can directly calculate the one-point Higgs tadpole diagrams, the two-point self-energies of physical particles, and the one-loop vertex corrections, in which the UV divergences are expected to be canceled out by the introduced counterterms. The dimensional regularization is utilized to calculate the loop functions and separate the divergent terms from finite ones~\cite{Bollini:1972ui,tHooft:1972tcz}, where the space-time dimension is set to $d = 4-2\epsilon$ and the UV divergence is denoted as $\Delta \equiv 1/\epsilon - \gamma_{\rm E}^{} + \ln(4\pi)$ with $\gamma_{\rm E}^{} \approx 0.577$ being the Euler-Mascheroni constant. To fix those counterterms, one has to impose the renormalization conditions. In the $\overline{\rm MS}$ scheme, the counterterms are determined to simply cancel out the UV-divergent terms from one-loop amplitudes proportional to $\Delta$. With the finite one-loop corrections, we are able to compute the UV-finite $S$-matrix elements for any physical processes. Since all renormalization schemes need to remove the UV divergences, the counterterms in the $\overline{\rm MS}$ scheme and their counterparts in the on-shell scheme should have the same divergent terms.\footnote{In fact, it is also possible to first fix the on-shell counterterms by the on-shell renormalization conditions, and identify the divergent parts as the $\overline{\rm MS}$ counterterms. In this way, the matching conditions between the on-shell parameters and the $\overline{\rm MS}$ ones at the one-loop order can be obtained.} 
	\end{itemize}
	
	Within the above renormalization scheme, we need to further compute all the one-particle irreducible (1PI) two-point functions for gauge bosons, the Higgs boson and fermions to determine the renormalization constants of wave functions. In the type-I seesaw model, except for the lepton self-energies, only the fermionic contributions to the self-energies of gauge and scalar bosons are modified by heavy Majorana neutrinos, while other corrections, such as the quark self-energies and the bosonic contributions to gauge bosons and scalars are the same as those in the SM~\cite{Marciano:1980pb,Denner:1991kt,Huang:2023nqf,Degrassi:1992ff}. Therefore, we shall focus on the leptonic sector in the following, especially the counterterms of lepton masses, wave functions and the flavor mixing matrix. For simplicity, the scale-dependence of physical parameters will not be explicitly shown, but we should always keep in mind that those are running parameters.
	
\section{Renormalization-group Equations}
	
	\label{sec:RGE}
	
	\subsection{Fermion self-energies and counterterms}
	
	We start our discussions with the self-energies of Dirac fermion fields $f$, which can be separated into the left- and right-handed fields as $f = f^{\rm L}_{} + f^{\rm R}_{}$. The bare Lagrangian of a Dirac fermion $f$ including the kinetic and mass terms reads
	\begin{eqnarray}
		{\cal L}_0^{\rm f} = \overline{f^{\rm L}_{0}} \left({\rm i} \slashed{\partial}\right) f^{\rm L}_{0} + \overline{f^{\rm R}_{0}} \left({\rm i} \slashed{\partial}\right) f^{\rm R}_{0} - \overline{f^{\rm L}_{0}} m_{f,0}^{} f^{\rm R}_{0} - \overline{f^{\rm R}_{0}} m_{f,0}^{} f^{\rm L}_{0} \;.
	\end{eqnarray}
	With the renormalized masses and fields defined in Eqs.~(\ref{eq:ParaCT}) and (\ref{eq:WavFuncCT}), ${\cal L}_0^{\rm f}$ can be rewritten in terms of the renormalized one
	\begin{eqnarray}
		{\cal L}_{}^{\rm f} = \overline{f} {\rm i} \slashed{\partial} f - \overline{f} m_f^{} f \; ,
	\end{eqnarray}
	and the corresponding counterterm
	\begin{eqnarray}
		\delta {\cal L}_{}^{\rm f} &=& \overline{f_i^{}} \left[\frac{1}{2} \left(\delta Z_{ij}^{\rm L\dagger} + \delta Z_{ij}^{\rm L}\right) {\rm i} \slashed{\partial} P_{\rm L}^{} + \frac{1}{2} \left(\delta Z_{ij}^{\rm R\dagger} + \delta Z_{ij}^{\rm R}\right) {\rm i} \slashed{\partial} P_{\rm R}^{} \right. \nonumber \\
		&& \left. - \left(\frac{1}{2} \delta Z_{ij}^{\rm R\dagger} m_j^{} + \frac{1}{2} m_i^{} \delta Z_{ij}^{\rm L} + \delta_{ij}^{} \delta m_i^{}\right) P_{\rm L}^{} - \left(\frac{1}{2}\delta Z_{ij}^{\rm L\dagger} m_j^{} + \frac{1}{2} m_i^{} \delta Z_{ij}^{\rm R} + \delta_{ij}^{} \delta m_i^{}\right) P_{\rm R}^{} \right] f_j^{} \;. \qquad 
	\end{eqnarray}
The Feynman rule for the fermion two-point counterterm is
	\begin{eqnarray}
		\label{eq:fermion_prop_ct}
		\begin{tikzpicture}[baseline=0]
			\begin{feynman}
				% 定义顶点
				\vertex (a) {$f_j^{}$};
				\vertex [right=1.5cm of a] (c);
				\vertex [right=3cm of a] (b) {$f_i^{}$};
				
				% 绘制费曼图并标注动量
				\diagram* {
					(a) -- [momentum= $p$, insertion={0.5}] (b) ,
					(a) --[fermion] (c),
					(c) --[fermion] (b)
				};
			\end{feynman}
		\end{tikzpicture} = {\rm i} \left(C_{\rm L}^{} \slashed{p} P_{\rm L}^{} + C_{\rm R}^{} \slashed{p} P_{\rm R}^{} - C_{\rm l}^{} P_{\rm L}^{} - C_{\rm r}^{} P_{\rm R}^{}\right) \;,
	\end{eqnarray}
	with
	\begin{eqnarray}
		C_{\rm L}^{} &=& \frac{1}{2} \left(\delta Z_{ij}^{\rm L\dagger} + \delta Z_{ij}^{\rm L}\right) \;, \nonumber \\
		C_{\rm R}^{} &=& \frac{1}{2} \left(\delta Z_{ij}^{\rm R\dagger} + \delta Z_{ij}^{\rm R}\right) \;, \nonumber \\
		C_{\rm l}^{} &=& \frac{1}{2} \delta Z_{ij}^{\rm R\dagger} m_j^{} + \frac{1}{2} m_i^{} \delta Z_{ij}^{\rm L} + \delta_{ij}^{} \delta m_i^{} \;, \nonumber \\
		C_{\rm r}^{} &=& \frac{1}{2}\delta Z_{ij}^{\rm L\dagger} m_j^{} + \frac{1}{2} m_i^{} \delta Z_{ij}^{\rm R} + \delta_{ij}^{} \delta m_i^{} \;. 
	\end{eqnarray}
	
The most general form of the two-point self-energy $\Sigma_{ij}^{} (p)$ for the Dirac field $f$ can always be decomposed into~\cite{Kniehl:1996bd}
	\begin{eqnarray}
		\label{eq:SE_fermion}
		\Sigma_{ij}^{} (p) \equiv \slashed{p} P_{\rm L}^{} \Sigma_{ij}^{\rm L}(p^2) + \slashed{p} P_{\rm R}^{} \Sigma_{ij}^{\rm R}(p^2) + P_{\rm L}^{} \Sigma_{ij}^{\rm D}(p^2) + P_{\rm R}^{} \Sigma_{ij}^{\rm D\dagger}(p^2) \;,
	\end{eqnarray}
where the superscripts ``L", ``R", ``D" refer to the left-handed, right-handed and scalar parts. The hermiticity of the effective action requires the left- and right-handed parts satisfying $\Sigma_{ij}^{\rm L,R}(p^2) = \Sigma_{ji}^{\rm L,R*}(p^2)$. Given the counterterm and the unrenormalized self-energy in Eqs.~(\ref{eq:fermion_prop_ct})-(\ref{eq:SE_fermion}), the renormalized self-energy can be expressed as
	\begin{eqnarray}
		\label{eq:renorm_SE}
		\widehat{\Sigma}_{ij}^{\rm L} (p^2) &=& \Sigma_{ij}^{\rm L} (p^2) + \frac{1}{2} \left(\delta Z_{ij}^{\rm L\dagger} + \delta Z_{ij}^{\rm L}\right) \;, \nonumber \\
		\widehat{\Sigma}_{ij}^{\rm R} (p^2) &=& \Sigma_{ij}^{\rm R} (p^2) + \frac{1}{2} \left(\delta Z_{ij}^{\rm R\dagger} + \delta Z_{ij}^{\rm R}\right) \;, \nonumber \\
		\widehat{\Sigma}_{ij}^{\rm D} (p^2) &=& \Sigma_{ij}^{\rm D} (p^2) - \left(\frac{1}{2} \delta Z_{ij}^{\rm R\dagger} m_j^{} + \frac{1}{2} m_i^{} \delta Z_{ij}^{\rm L} + \delta_{ij}^{} \delta m_i^{}\right) \;.
	\end{eqnarray}
Then, the wave-function counterterms are fixed by imposing the $\overline{\rm MS}$ renormalization conditions, namely, absorbing the UV-divergent terms of self-energies into counterterms and making the renormalized self-energies UV-finite. In the case of $i \neq j$, the wave-function counterterms can be found from Eq.~(\ref{eq:renorm_SE}) as
	\begin{eqnarray}
		\delta Z_{ij}^{\rm L} &=& \frac{2}{m_i^2 - m_j^2} \left(m_j^2 \Sigma_{ij}^{\rm L} + m_i^{} m_j^{} \Sigma_{ij}^{\rm R} + m_i^{} \Sigma_{ij}^{\rm D} + m_j^{} \Sigma_{ij}^{\rm D\dagger} \right)_{\rm div}^{} \;, \nonumber \\
		\delta Z_{ij}^{\rm R} &=& \frac{2}{m_i^2 - m_j^2} \left(m_j^2 \Sigma_{ij}^{\rm R} + m_i^{} m_j^{} \Sigma_{ij}^{\rm L} + m_j^{} \Sigma_{ij}^{\rm D} + m_i^{} \Sigma_{ij}^{\rm D\dagger}\right)_{\rm div}^{} \; ,
	\end{eqnarray}
	where the subscript ``div" refers to the UV-divergent parts in the one-loop self-energies, i.e., terms proportional to $\Delta$. Similarly, for $i=j$, one gets the mass counterterm
	\begin{eqnarray}
		\delta m_i^{} = \frac{1}{2} m_i^{} \left(\Sigma_{ii}^{\rm L} + \Sigma_{ii}^{\rm R}\right)_{\rm div}^{} + \frac{1}{2} \left(\Sigma_{ii}^{\rm D} + \Sigma_{ii}^{\rm D *}\right)_{\rm div}^{} \;,
	\end{eqnarray}
	and
	\begin{eqnarray}
		{\rm Re}\,\delta Z_{ii}^{\rm L} = - \Sigma_{ii,{\rm div}}^{\rm L} \;, \quad {\rm Re}\,\delta Z_{ii}^{\rm R} = - \Sigma_{ii,{\rm div}}^{\rm R} \;.
	\end{eqnarray}
On the other hand, the imaginary parts of the wave-function counterterms should fulfill the relation
	\begin{eqnarray}
		\label{eq:Im_delta_Z}
		m_i^{} {\rm Im} \left(\delta Z_{ii}^{\rm L} - \delta Z_{ii}^{\rm R}\right) = 2 {\rm Im}\,\Sigma_{ii,{\rm div}}^{\rm D} \;.
	\end{eqnarray}
This relation implies that the diagonal wave-function counterterms cannot be uniquely determined due to the rephasing symmetry of Dirac fermion fields~\cite{Kniehl:1996bd}.
	
All discussions above can be applied to the case of Majorana fermions. However, due to the Majorana conditions, we have $f^{\rm L,R}_{0} = \left(f^{\rm R,L}_{\rm 0}\right)^{\rm c}$ for bare fields and $f^{\rm L,R}_{} = \left(f^{\rm R,L}_{}\right)^{\rm c} $ for renormalized fields, indicating $\delta Z^{\rm L}_{} = \delta Z_{}^{\rm R*}$. Furthermore, the left- and right-handed parts of the self-energies are now related by $\Sigma_{ij}^{\rm L,R} = \Sigma_{ji}^{\rm R,L} = \Sigma_{ij}^{\rm R,L*}$, and the relation $\Sigma_{ij}^{\rm M} = \Sigma_{ji}^{\rm M}$ holds for the scalar part.\footnote{A detailed proof of these relations can be found in Ref.~\cite{Grimus:2016hmw}.} The corresponding wave-function counterterms are
	\begin{eqnarray}
		\delta Z_{ij}^{\rm L} &=& \delta Z_{ij}^{\rm R*} = \frac{2}{m_i^2 - m_j^2} \left(m_j^2 \Sigma_{ij}^{\rm L} + m_i^{} m_j^{} \Sigma_{ij}^{\rm R} + m_i^{} \Sigma_{ij}^{\rm M} + m_j^{} \Sigma_{ij}^{\rm M\dagger} \right)_{\rm div}^{} \;,
	\end{eqnarray}
for $i\neq j$. In the diagonal case with $i=j$, we have the mass counterterms
	\begin{eqnarray}
		\delta m_i^{} = m_i^{} \Sigma_{ii,{\rm div}}^{\rm L} + \frac{1}{2} \left(\Sigma_{ii}^{\rm M} + \Sigma_{ii}^{\rm M *}\right)_{\rm div}^{} \;.
	\end{eqnarray}
and the wave-function counterterms
	\begin{eqnarray}
		\delta Z_{ii}^{\rm L} = - \Sigma_{ii,{\rm div}}^{\rm L} + \frac{1}{2 m_i^{}} \left(\Sigma_{ii}^{\rm M} - \Sigma_{ii}^{\rm M *}\right)_{\rm div}^{} \;,
	\end{eqnarray}
where it should be noticed that no rephasing symmetry exists in the Majorana case. In the SM, it is possible to further decompose the scalar part into $\Sigma_{ij}^{\rm D} = m_i^{} \Sigma_{ij}^{\rm S}$ with $\Sigma_{ij}^{\rm S}$ being a Hermitian matrix. However, such a decomposition is inapplicable to Majorana fermions~\cite{Kniehl:1996bd}. 
	
Given the wave-function counterterms of lepton fields, it is straightforward to figure out the counterterm of the flavor mixing matrix ${\cal U}$. Only the elements in the first three rows of ${\cal U}$, i.e., those of ${\cal B}$, appear in the interactions and are physical, whose counterterms are fixed through~\cite{Denner:1990yz,Kniehl:1996bd}
	\begin{eqnarray}
		\delta {\cal B}_{\alpha i}^{} = - \frac{1}{4} \left[\sum_{k=1}^6 {\cal B}_{\alpha k}^{} \left(\delta Z_{ki}^{\rm \chi,L} - \delta Z_{ki}^{\rm \chi,L \dagger}\right) + \sum_{\beta=e,\mu,\tau} \left(\delta Z_{\alpha \beta}^{l,{\rm L}\dagger} - \delta Z_{\alpha \beta}^{l,{\rm L}}\right) {\cal B}_{\beta i}^{}\right] \;.
	\end{eqnarray}
	It is clear that only the anti-Hermitian parts of the wave-function counterterms contribute. In addition, with ${\cal C} \equiv {\cal B}^\dagger {\cal B}$, its counterterm is obtained directly as $\delta {\cal C}_{ij}^{} = \delta{\cal B}^\dagger_{i \alpha} {\cal B}_{\alpha j}^{} + {\cal B}^\dagger_{i\alpha} \delta{\cal B}_{\alpha j}^{}$. 
	
The full expressions of the two-point functions for charged leptons $\Sigma_{\alpha \beta}^{}(p^2)$ and Majorana neutrinos $\Sigma_{ij}^{}(p^2)$ in the $R_\xi^{}$ gauge are given in Appendix~\ref{app:1-loop}, where the Passarino-Veltman (PV) functions are implemented for the loop integrals~\cite{Passarino:1978jh}. Tadpole contributions are included so that the renormalized masses and the mass counterterms are both gauge-independent. With the help of these self-energies, we may explicitly calculate the wave-function and mass counterterms for neutrinos and charged leptons. 
\begin{itemize}
  \item First, for Majorana neutrinos, the left-handed wave-function counterterms are 	
	\begin{eqnarray}
		\delta Z_{ij}^{\rm \chi,L} &=& \frac{g^2 \Delta}{2 (4\pi)^2 m_W^2 \left( \widehat{m}_i^2 - \widehat{m}_j^2 \right)} \left\{ - \sum_\alpha m_\alpha^2 \left[3 {\cal B}_{\alpha i}^{} {\cal B}_{\alpha j}^{*} \widehat{m}_i^{} \widehat{m}_j^{} + {\cal B}_{\alpha i}^{*} {\cal B}_{\alpha j}^{} \left(2 \widehat{m}_i^2+\widehat{m}_j^2\right)\right] \right. \nonumber \\
		&& +{\cal C}_{ij}^{}  \left[ 2 \widehat{m}_i^2 \widehat{m}_j^2 -\left(\xi_W^{}+\frac{\xi_Z^{}}{2 c^2}\right) m_W^2 \left( \widehat{m}_i^2 - \widehat{m}_j^2 \right) \right] + 2 {\cal C}_{ij}^{*} \widehat{m}_i^{} \widehat{m}_j^3  \nonumber \\
		&& \vphantom{\left[\sum_k\right]} + \sum_k m_k^2 \left({\cal C}_{ik}^{} {\cal C}_{kj}^{} \widehat{m}_j^{2} + {\cal C}_{ik}^{*} {\cal C}_{kj}^{*} \widehat{m}_i^{} \widehat{m}_j^{} \right) + \left[{\cal C}_{ij}^{} \left(\widehat{m}_i^2+\widehat{m}_j^2\right)+2 {\cal C}_{ij}^{*} \widehat{m}_i^{} \widehat{m}_j^{}\right]  \nonumber \\
		&& \left. \times \left(-3 \frac{m_h^4 + 4 m_W^4 +2 m_Z^4}{2m_h^2} + 4 \sum_{f=q,l} \frac{m_f^4}{m_h^2} + 4 \sum_{k} {\cal C}^{}_{kk} \frac{\widehat{m}_k^4}{m_h^2} \right) \right\} 
	\end{eqnarray}
	for $i\neq j$, where the summation over charged-fermion species (i.e., quarks $q$ and charged leptons $l$) and all Majorana neutrinos $\chi_k^{}$ in the last term are expressed separately. They come from the tadpole diagrams, and the color factor of three should be multiplied for quarks. The right-handed counterterms satisfy $\delta Z_{ij}^{\rm \chi,R} = \delta Z_{ij}^{\rm \chi,L*}$. For $i=j$, the counterterms of wave functions are
	\begin{eqnarray}
		\delta Z_{ii}^{\rm \chi,L} = - \frac{g^2 \Delta}{4 (4\pi)^2} \left[ \sum_\alpha \left|{\cal B}_{\alpha i}^{}\right|^2 \frac{m_\alpha^2}{m_W^2} + {\cal C}_{ii}^{} \left(\frac{\xi_Z^{}}{c^2} + 2 \xi_W^{} + \frac{2 \widehat{m}_i^2}{m_W^2}\right) + \sum_k \left|{\cal C}_{ik}^{}\right|^2 \frac{\widehat{m}_k^2}{m_W^2} \right] \;,
	\end{eqnarray}
	which are real and do not contribute to $\delta{\cal B}$, and the neutrino mass counterterms are
	\begin{eqnarray}
		\delta \widehat{m}_i^{} &=& \frac{g^2 \widehat{m}_i^{} \Delta}{4 (4\pi)^2} \left(- 3 \sum_\alpha \left|{\cal B}_{\alpha i}^{}\right|^2 \frac{m_\alpha^2}{m_W^2} + 2 {\cal C}_{ii}^{} \frac{\widehat{m}_i^2}{m_W^2} + \sum_k \left|{\cal C}_{ik}^{}\right|^2 \frac{\widehat{m}_k^2}{m_W^2}   \right. \nonumber \\
		&& \left. - 3 {\cal C}_{ii}^{} \frac{m_h^4 + 4 m_W^4 + 2 m_Z^4 }{m_h^2 m_W^2} + 8 {\cal C}_{ii}^{} \sum_{f=q,l} \frac{m_f^4}{m_h^2 m_W^2} + 8 {\cal C}_{ii}^{}  \sum_{k} {\cal C}_{kk}^{} \frac{\widehat{m}_k^4}{m_h^2 m_W^2}  \right)\;.
	\end{eqnarray}
	It is gauge-independent with tadpole contributions in the second line, and can be used to calculate the RGEs of Majorana neutrino masses.
	
\item	For charged leptons, the off-diagonal counterterms of wave functions with $\alpha\neq \beta$ receive only the contributions from massive neutrinos, i.e.,
	\begin{eqnarray}
		\delta Z_{\alpha \beta}^{l,{\rm L}} &=& - \frac{g^2 \Delta}{2 (4\pi)^2} \frac{2 m_\alpha^2 + m_\beta^2}{m_\alpha^2 - m_\beta^2} \sum_i {\cal B}_{\alpha i}^{} {\cal B}_{\beta i}^{*} \frac{\widehat{m}_i^2}{m_W^2}  \;, \nonumber \\
		\delta Z_{\alpha \beta}^{l,{\rm R}} &=& -\frac{3  g^2 \Delta}{2 (4\pi)^2} \frac{m_\alpha^{} m_\beta^{}}{m_\alpha^2 - m_\beta^2} \sum_i {\cal B}_{\alpha i}^{} {\cal B}_{\beta i}^{*} \frac{\widehat{m}_i^2}{m_W^2} \;, 
	\end{eqnarray}
	while the diagonal wave-function and mass counterterms are
	\begin{eqnarray}
		\delta Z_{\alpha\alpha}^{l,{\rm L}} &=& -\frac{g^2 \Delta}{4 (4\pi)^2 } \left(\sum_i \left|{\cal B}_{\alpha i}^{}\right|^2 \frac{\widehat{m}_i^2}{m_W^2} + \frac{m_\alpha^2}{m_W^2} + 4 \xi_A^{} s^2 + 2 \xi_W^{} + \frac{4 c_{\rm L}^2}{c^2} \xi_Z^{}\right) \;, \\
		\delta Z_{\alpha\alpha}^{l,{\rm R}} &=& -\frac{g^2 \Delta}{4 (4\pi)^2} \left(\frac{2 m_\alpha^2}{m_W^2} + \frac{4 c_{\rm R}^2 }{c^2} \xi_Z^{} + 4 \xi_A^{}  s^2\right) \;, \\
		\delta m_\alpha^{} &=& - \frac{3 g^2 m_\alpha^{} \Delta}{8 (4\pi)^2 } \left( \sum_i \left|{\cal B}_{\alpha i}^{}\right|^2 \frac{\widehat{m}_i^2}{m_W^2} +  \frac{4 m_Z^2 - 4 m_W^2 - m_\alpha^2}{m_W^2} \right. \nonumber \\
		&& \left. + \frac{2 m_Z^4 + 4 m_W^4 + m_h^4}{m_h^2 m_W^2} - \frac{8}{3} \sum_{f=q,l} \frac{m_f^4}{m_h^2 m_W^2} - \frac{8}{3} \sum_k {\cal C}_{kk}^{} \frac{\widehat{m}_k^4}{m_h^2 m_W^2} \right) \;.
	\end{eqnarray}
Note that the values of the left- and right-handed couplings $c_{\rm L}^{} = -1/2 + s^2$ and $c_{\rm R}^{} = s^2$ have already been input into the last equation to demonstrate the gauge-independence of mass counterterms.
\end{itemize}
	
\subsection{RGEs for Physical Parameters}
	
With the $\overline{\rm MS}$ counterterms for masses and mixing matrices given in the previous subsection, one can derive the RGEs for physical parameters at the one-loop level. For the running parameter $a(\mu)$ with its $\overline{\rm MS}$ counterterm $\delta a$, since the bare parameter $a_0^{}$ is independent of the renormalization scale $\mu$, its one-loop RGE can be calculated as~\cite{tHooft:1973mfk}
	\begin{eqnarray}\label{eq:adot}
		\dot{a} \equiv \frac{{\rm d} a}{{\rm d} t} = - \frac{{\rm d} \delta a}{{\rm d} t} = \lim_{\epsilon \to 0} \epsilon g \frac{\partial \delta a}{\partial g} + {\cal O}(g^3) \;,
	\end{eqnarray}
where $t \equiv \ln\left(\mu/\Lambda_{\rm R}^{}\right)$ with $\mu$ being an arbitrary renormalization scale, and the fact that ${\rm d}g/{\rm d}t = - \epsilon g + {\cal O}(g^3)$ should be noticed. Following the approach in Eq.~(\ref{eq:adot}), we arrive at the RGEs for neutrino masses
	\begin{eqnarray}
		16\pi^2 \dot{\widehat{m}}_i^{} &=& \frac{g^2}{2} \widehat{m}_i^{} \left( -3 \sum_\alpha \left|{\cal B}^{}_{\alpha i}\right|^2 \frac{m_\alpha^2}{m_W^2} + 2 {\cal C}^{}_{ii} \frac{\widehat{m}_i^2}{m_W^2} + \sum_k \left|{\cal C}^{}_{ik}\right|^2 \frac{\widehat{m}_k^2}{m_W^2} \right. \nonumber \\
		&& \left. - 3 {\cal C}^{}_{ii} \frac{m_h^4 + 4 m_W^4 + 2 m_Z^4}{m_h^2 m_W^2} + 8 {\cal C}_{ii}^{} \sum_{f=q,l} \frac{m_f^4}{m_h^2 m_W^2} + 8 {\cal C}_{ii}^{}  \sum_{k} {\cal C}_{kk}^{} \frac{\widehat{m}_k^4}{m_h^2 m_W^2} \right) \;, 
	\end{eqnarray}
and
	\begin{eqnarray}
		16\pi^2  \dot{m}_\alpha^{} &=& \frac{g^2}{2} m_\alpha^{} \left(-\frac{3}{2} \sum_i \left|{\cal B}^{}_{\alpha i}\right|^2 \frac{\widehat{m}_i^2}{m_W^2} - 6 \frac{m_Z^2}{m_W^2} + \frac{3 m_\alpha^2}{2 m_W^2}+6 \right. \nonumber \\
		&& \left.  - 3 \frac{m_h^4 + 4 m_W^4 + 2 m_Z^4}{2 m_h^2 m_W^2} + 4 \sum_{f=q,l} \frac{m_f^4}{m_h^2 m_W^2} + 4 \sum_k {\cal C}_{kk}^{} \frac{\widehat{m}_k^4}{m_h^2 m_W^2} \right) \;,
	\end{eqnarray}
for charged-lepton masses. To the best of our knowledge, the above RGEs for neutrino masses and charged-lepton masses in the full type-I seesaw model have not been explicitly calculated in the literature before.

With the off-diagonal wave-function counterterms of neutrinos and charged leptons, the RG running of the leptonic mixing matrix elements ${\cal B}_{\alpha i}^{}$ is governed by
	\begin{eqnarray}
		\label{eq:RG_B}
		16\pi^2 \dot{{\cal B}}_{\alpha i}^{} &=& \frac{g^2}{4} \left\{\sum_{\beta \neq \alpha} \frac{m_\alpha^2 + m_\beta^2}{m_\beta^2 - m_\alpha^2}  \sum_j {\cal B}_{\alpha j}^{} {\cal B}_{\beta j}^* {\cal B}_{\beta i}^{} \frac{3 \widehat{m}_j^2}{m_W^2} - \sum_{k\neq i} \frac{{\cal B}_{\alpha k}^{}}{\widehat{m}_k^2 - \widehat{m}_i^2} \right. \nonumber\\
		&&  \times \left[\left(\widehat{m}_i^2 + \widehat{m}_k^2\right) \left( 2 {\cal C}_{ki}^* \frac{\widehat{m}_i^{} \widehat{m}_k^{}}{m_W^2} + \sum_n {\cal C}_{kn}^{} {\cal C}_{ni}^{} \frac{\widehat{m}_n^2}{m_W^2} - \sum_\rho {\cal B}_{\rho i}^{} {\cal B}_{\rho k}^* \frac{3 m_\rho^2}{m_W^2} \right) \right. \nonumber \\
		&& + 2 \widehat{m}_i^{} \widehat{m}_k^{} \left( 2 {\cal C}_{ki}^{} \frac{\widehat{m}_i^{} \widehat{m}_k^{}}{m_W^2} + \sum_n {\cal C}_{kn}^* {\cal C}_{ni}^* \frac{\widehat{m}_n^2}{m_W^2} - \sum_\rho {\cal B}_{\rho i}^* {\cal B}_{\rho k}^{} \frac{3 m_\rho^2}{m_W^2} \right) \nonumber \\
		&& - 3 \left(2 {\cal C}_{ki}^* \frac{\widehat{m}_i^{} \widehat{m}_k^{}}{m_W^2} + {\cal C}_{ki}^{} \frac{\widehat{m}_i^2 + \widehat{m}_k^2}{m_W^2}\right)  \nonumber \\
		&& \left. \left. \times \left(\frac{m_h^4 + 4 m_W^4 + 2 m_Z^4}{m_h^2} - \frac{8}{3} \sum_{f=q,l} \frac{m_f^4}{m_h^2} - \frac{8}{3} \sum_j \frac{{\cal C}_{jj}^{} \widehat{m}_j^4}{m_h^2} \right)\right] \right\} \;,
	\end{eqnarray}
where the factor of two in front of ${\cal C}_{ki}^*$ and ${\cal C}_{ki}^{}$ in the second and third lines has been corrected in comparison with the existing results in Ref.~\cite{Pilaftsis:2002nc}, and new terms in the last two lines come from the tadpole diagrams with fermions and the Higgs boson running in the loop. Except for these contributions, we have checked that our results Eq.~(\ref{eq:RG_B}) are consistent with those in Ref.~\cite{Pilaftsis:2002nc}.
	
For completeness, we provide the RGEs for the fine-structure constant, gauge-boson masses, the Higgs-boson mass, quark masses and the CKM matrix elements, in order to ensure that the RGEs are complete and closed. Most of these RGEs are the same as in the SM, but the leptonic contributions due to the presence of massive Majorana neutrinos are new. With the electric charge $Q_f^{}$ of the fermion $f$, the beta function for the fine-structure constant reads
\begin{eqnarray}
	\dot{\alpha} = \frac{\alpha^2}{\pi} \left(- \frac{7}{2} + \frac{2}{3} \sum_{f=q,l} Q_f^2 \right) \;.
\end{eqnarray}
We recalculate two-point functions of weak gauge bosons and the Higgs boson in the $R_\xi^{}$ gauge. The renormalization conditions of their mass counterterms are
	\begin{eqnarray}
		\delta m_{W}^2 = - \Sigma_{\rm T,div}^{W} (m_{W}^{2}) \;, \quad \delta m_{Z}^2 = - \Sigma_{\rm T,div}^{Z} (m_{Z}^{2}) \;, \quad \delta m_{h}^2 = + \Sigma_{\rm div}^{h} (m_{h}^{2}) \;,
	\end{eqnarray}
	where $\Sigma_{\rm T,div}^{W,Z}$ and $\Sigma_{\rm div}^{h}$ denote the UV-divergent terms of the transverse self-energies of weak gauge bosons and the Higgs boson, respectively. Their RGEs can be obtained through
	\begin{eqnarray}
		\dot{m}_X^{} = \lim_{\epsilon \to 0} \frac{\epsilon g}{2 m_X^{}} \frac{\partial \delta m_X^2}{\partial g} + {\cal O}(g^3)\;,
	\end{eqnarray}
	for $X=W,Z,h$. More explicitly, for the $Z$-boson mass, we have
	\begin{eqnarray}
		16\pi^2 \dot{m}_Z^{} &=& - g^2 m_Z^{} \left\{ \frac{84 c^4-14 c^2-11}{12 c^2} + 3 \frac{m_h^4 + 4 m_W^4 + 2 m_Z^4 }{4 m_W^2 m_h^2} \right. \nonumber \\
		&& - \sum_{f=q,l} \left[\frac{\left(c_{\rm A}^{f}\right)^2+\left(c_{\rm V}^f\right)^2}{3 c^2}-\frac{2 \left(c_{\rm A}^{f}\right)^2 m_f^2}{m_W^2} + \frac{2  m_f^4}{m_W^2 m_h^2}\right] \nonumber \\
		&& \left. + \sum_{i,j} \frac{3 \left({\cal C}_{ij}^2 + {\cal C}_{ij}^{*2} \right) \widehat{m}_i^{} \widehat{m}_j^{} + \left|{\cal C}_{ij}^{}\right|^2  \left(3 \widehat{m}_i^2 + 3 \widehat{m}_j^2 - 2 m_Z^2\right)}{12 m_W^2 } - 2 \sum_k \frac{{\cal C}_{kk}^{} \widehat{m}_k^4}{m_h^2 m_W^2} \right\} \;,
	\end{eqnarray}
	where $c_{\rm V}^f \equiv I_3^f - 2 s^2 Q_f^{}$ and $c_{\rm A}^f \equiv I_3^f$ refer respectively to the vector-type and axial-vector-type couplings of the NC interaction, with $I_3^f$ being the weak isospin component of $f$. In a similar way, the RG running of the $W$-boson mass is determined by
	\begin{eqnarray}
		16\pi^2 \dot{m}_W^{} &=& - g^2 m_W^{} \left( \frac{68c^2-9}{12 c^2} + 3 \frac{m_h^4 + 4 m_W^4 + 2 m_Z^4}{4 m_h^2 m_W^2} \right. \nonumber \\
		&& + \sum_{\left\{q,q'\right\}} \left|{\bf V}_{q q'}^{\rm CKM}\right|^2 \frac{3 m_q^2 + 3 m_{q'}^2 - 2 m_W^2}{6 m_W^2} + \sum_{\left\{l_\alpha^{}, \chi_i^{}\right\}} \left|{\cal B}_{\alpha i}^{}\right|^2 \frac{ 3 m_\alpha^2+3 \widehat{m}_i^2-2 m_W^2}{6 m_W^2 }   \nonumber \\
		&& \left. - 2 \sum_{f=q,l}\frac{m_f^4}{m_h^2 m_W^2 }   -  2 \sum_k \frac{{\cal C}_{kk}^{} \widehat{m}_k^4}{m_h^2 m_W^2 } \right) \;,
	\end{eqnarray}
	with $\left\{q,q'\right\}$ denoting all pairs of up- and down-type quarks. Finally, the RGE for the Higgs-boson mass is
	\begin{eqnarray}
		16\pi^2 \dot{m}_h^{} &=& g^2 m_h^{} \left[3 \frac{m_h^2 - 2 m_W^2 - m_Z^2}{4 m_W^2} + \sum_{f=q,l} \frac{m_f^2}{2 m_W^2} + \sum_k 3 \frac{{\cal C}_{kk}^{} \widehat{m}_k^4}{m_W^2 m_h^2 } \right. \nonumber \\
		&& + \sum_{i,j} \left({\cal C}_{ij}^2 + {\cal C}_{ij}^{*2} \right) \frac{  \widehat{m}_i^{} \widehat{m}_j^{} \left(m_h^2-3 \widehat{m}_i^2-3\widehat{m}_j^2\right)}{4 m_W^2 m_h^2 }  \nonumber \\
		&& \left. + \left|{\cal C}_{ij}^{}\right|^2  \frac{m_h^2 \left(\widehat{m}_i^2+\widehat{m}_j^2\right)-2 \left(\widehat{m}_i^4+4 \widehat{m}_i^2 \widehat{m}_j^2+\widehat{m}_j^4\right)}{4 m_W^2 m_h^2 } \right] \;.
	\end{eqnarray}
	
	In the quark sector, since no contributions from massive neutrinos exist at the one-loop level, the RGEs for quark masses are the same as in the SM. For the up-type quark masses, we have
	\begin{eqnarray}
		16\pi^2 \dot{m}_q^{} &=& \frac{g^2}{4} m_q^{} \left( 3 \frac{m_q^2 - 8 s^2 m_Z^2 Q_q Y_q}{ m_W^2} - 3 \sum_{q'} \left|{\bf V}_{q q'}^{\rm CKM}\right|^2 \frac{m_{q'}^2}{m_W^2} - 3 \frac{m_h^4 + 4 m_W^4 + 2 m_Z^4}{m_h^2 m_W^2 } \right. \nonumber \\
		&& \left. + 8 \sum_{f=q,l} \frac{m_f^4}{m_h^2 m_W^2} + 8 \sum_k \frac{{\cal C}_{kk}^{} \widehat{m}_k^2}{ m_h^2 m_W^2} \right) \;,
	\end{eqnarray}
	with $q'$ being the isospin partner of $q$, and the weak hypercharge $Y_q^{} = 1/6$ according to the Gell-Mann-Nishijima formula $Q_f^{} = I_3^f + Y_f^{}$~\cite{Nakano:1953zz,Gell-Mann:1956iqa}. For the down-type quarks, ${\bf V}_{q q'}^{\rm CKM}$ should be replaced by ${\bf V}_{q' q}^{\rm CKM}$. The $\overline{\rm MS}$ counterterm of the CKM matrix can be found in Ref.~\cite{Denner:1990yz}, while the corresponding beta functions for its matrix elements are given by
	\begin{eqnarray}
		16\pi^2 \dot{\bf V}_{ud}^{\rm CKM} &=& - g^2 \sum_{u'\neq u,d'\neq d} {\bf V}^{\rm CKM}_{u d'} {\bf V}^{\rm CKM *}_{u'd'} {\bf V}^{\rm CKM}_{u'd} \nonumber \\
		&& \times \left\{\frac{1}{m_{u'}^2 - m_u^2} \left[\left(m_u^2 + m_{u'}^2\right) \left(-\frac{3}{4} \frac{m_{d'}^2}{m_W^2} + \frac{1}{2}\right) + \frac{m_u^2 m_{u'}^2}{2 m_W^2} \right] \right. \nonumber \\
		&& \left. + \left(u\leftrightarrow d, u' \leftrightarrow d'\right) \vphantom{\frac{1}{m_{u'}^2 - m_u^2} \left[\left(m_u^2 + m_{u'}^2\right) \left(-\frac{3}{4} \frac{m_{d'}^2}{m_W^2} + \frac{1}{2}\right) + \frac{m_u^2 m_{u'}^2}{2 m_W^2} \right]} \right\} \;,
	\end{eqnarray}
where $\left\{u,u'\right\}$ represent the up-type quarks and $\left\{d,d'\right\}$ the down-type quarks, respectively. 

Now we have the whole set of RGEs that are complete and self-consistent. One may choose one set of physical parameters at a specific energy scale as input and numerically solve this set of RGEs to get their values at another energy scale.
	
\subsection{Euler-like Parametrization}
	
As mentioned before, the $6\times 6$ unitary matrix ${\cal U}$ describes leptonic flavor mixing, but only two $3\times 3$ submatrices ${\bf V}$ and ${\bf R}$ are present in neutrino interactions. To specify the independent physical parameters for leptonic flavor mixing, one always adopts a particular parametrization of ${\cal U}$ including physical rotation angles and CP-violating phases. In this subsection, based on the RGEs of ${\cal B}_{\alpha i}^{}$ in Eq.~(\ref{eq:RG_B}), we examine the RG running of mixing angles and CP-violating phases in the Euler-like parametrization~\cite{Xing:2007zj,Xing:2011ur}. In this parametrization, the $6\times6$ unitary matrix ${\cal U}$ is written as
	\begin{eqnarray}
		\label{eq:EulerPara}
		{\cal U} = 
		\begin{pmatrix}
			{\bf 1} & {\bf 0} \\ {\bf 0} & {\bf U}_{0}^{}
		\end{pmatrix} 
		\begin{pmatrix}
			{\bf A} & {\bf R} \\ {\bf D} & {\bf B}
		\end{pmatrix}
		\begin{pmatrix}
			{\bf V}_0^{} & {\bf 0} \\ {\bf 0} & {\bf 1}
		\end{pmatrix} \;,
	\end{eqnarray}
where ${\bf V}^{}_0$ and ${\bf U}_0^{}$ are $3 \times 3$ unitary matrices describing the flavor mixing among light and heavy neutrinos, respectively, while the $3\times3$ matrices ${\bf A}, {\bf R}, {\bf D}$ and ${\bf B}$ characterize the interplay between light and heavy sectors. In comparison with the expression in Eq.~(\ref{eq:VRSU}), we notice that the $3\times3$ matrix ${\bf R}$ remains the same, and ${\bf V} = {\bf A} {\bf V}_0^{}$ is actually the Pontecorvo-Maki-Nakagawa-Sakata (PMNS) matrix~\cite{Pontecorvo:1957cp,Maki:1962mu,Pontecorvo:1967fh}. Additionally, we have ${\bf S} = {\bf U}_0^{} {\bf D} {\bf V}_0^{}$ and ${\bf U} = {\bf U}_0^{} {\bf B}$. 
	
More explicitly, ${\cal U}$ is parametrized by fifteen two-dimensional rotation matrices ${\cal O}_{ij}^{}$ (for $1 \leqslant i < j \leqslant6$), including fifteen rotation angles $\theta_{ij}^{}$ and fifteen phases $\delta_{ij}^{}$. For each rotation matrix ${\cal O}_{ij}^{}$, the diagonal $(i,i)$- and $(j,j)$-elements are set to $c_{ij}^{} \equiv \cos\theta_{ij}^{}$, while the off-diagonal $(i,j)$- and $(j,i)$-elements accordingly to $\hat{s}_{ij}^* \equiv {\rm e}^{-{\rm i} \delta_{ij}^{}} \sin\theta_{ij}^{} $ and $-\hat{s}_{ij}^{} \equiv - {\rm e}^{{\rm i} \delta_{ij}^{}} \sin\theta_{ij}^{} $. The other diagonal elements are set to one and the off-diagonal elements to zero. The product of these fifteen rotation matrices is arranged as~\cite{Xing:2007zj,Xing:2011ur}
\begin{eqnarray}
\begin{pmatrix}
			{\bf V}_0^{} & {\bf 0} \\ {\bf 0} & {\bf 1}
		\end{pmatrix} &=& {\cal O}_{23}^{} {\cal O}_{13}^{} {\cal O}_{12}^{} \;, \nonumber \\
\begin{pmatrix}
			{\bf A} & {\bf R} \\ {\bf D} & {\bf B}
		\end{pmatrix} &=& {\cal O}_{36}^{} {\cal O}_{26}^{} {\cal O}_{16}^{} {\cal O}_{35}^{} {\cal O}_{25}^{} {\cal O}_{15}^{} {\cal O}_{34}^{} {\cal O}_{24}^{} {\cal O}_{14}^{} \;,  \\
		\begin{pmatrix}
			{\bf 1} & {\bf 0} \\ {\bf 0} & {\bf U}_{0}^{}
		\end{pmatrix} &=& {\cal O}_{56}^{} {\cal O}_{46}^{} {\cal O}_{45}^{}  
\;.\nonumber
	\end{eqnarray}
The $3\times 3$ unitary matrix ${\bf V}_0^{}$, containing three mixing angles $\{\theta^{}_{12}, \theta^{}_{13}, \theta^{}_{23}\}$, turns out to be
	\begin{eqnarray}
		{\bf V}_0^{} &=& \begin{pmatrix}
			c^{}_{12} c^{}_{13} & \hat{s}^{*}_{12} c^{}_{13} & \hat{s}^{*}_{13} \\
			- \hat{s}^{}_{12} c^{}_{23} - c^{}_{12} \hat{s}^{}_{13} \hat{s}^{*}_{23} & c^{}_{12} c^{}_{23}-\hat{s}^{*}_{12} \hat{s}^{}_{13} \hat{s}^{*}_{23} & c^{}_{13} \hat{s}^{*}_{23} \\
			\hat{s}^{}_{12} \hat{s}^{}_{23} - c^{}_{12} \hat{s}^{}_{13} c^{}_{23} & - \hat{s}^{*}_{12} \hat{s}^{}_{13} c^{}_{23} - c^{}_{12} \hat{s}^{}_{23} & c^{}_{13} c^{}_{23} 
		\end{pmatrix} \;. 
	\end{eqnarray}
On the other hand, the $3\times 3$ matrix ${\bf A}$ takes on a lower-triangular form, which measures the unitarity violation of the PMNS matrix, i.e.,
	\begin{eqnarray}
		{\bf A} = \begin{pmatrix}
			c^{}_{14} c^{}_{15} c^{}_{16} & 0 & 0 \\
			\begin{array}{l}
				- c^{}_{14} \hat{s}^{}_{15} \hat{s}^{*}_{25} c^{}_{26} - c^{}_{14} c^{}_{15} \hat{s}^{}_{16} \hat{s}^{*}_{26} \\ - \hat{s}^{}_{14} \hat{s}^{*}_{24} c^{}_{25} c^{}_{26}
			\end{array} & c^{}_{24} c^{}_{25} c^{}_{26} & 0 \\
			\begin{array}{l}
				- c^{}_{14} \hat{s}^{}_{15} c^{}_{25} \hat{s}^{*}_{35} c^{}_{36} + c^{}_{14} \hat{s}^{}_{15} \hat{s}^{*}_{25} \hat{s}^{}_{26} \hat{s}^{*}_{36} \\
				- c^{}_{14} c^{}_{15} \hat{s}^{}_{16} c^{}_{26} \hat{s}^{*}_{36} + \hat{s}^{}_{14} \hat{s}^{*}_{24} \hat{s}^{}_{25} \hat{s}^{*}_{35} c^{}_{36} \\
				+ \hat{s}^{}_{14} \hat{s}^{*}_{24} c^{}_{25} \hat{s}^{}_{26} \hat{s}^{*}_{36} -  \hat{s}^{}_{14} c^{}_{24} \hat{s}^{*}_{34} c^{}_{35} c^{}_{36} 
			\end{array}
			& 
			\begin{array}{l}
				- c^{}_{24}  \hat{s}^{}_{25} \hat{s}^{*}_{35} c^{}_{36} - c^{}_{24} c^{}_{25} \hat{s}^{}_{26} \hat{s}^{*}_{36} \\
				- \hat{s}^{}_{24} \hat{s}^{*}_{34} c^{}_{35} c^{}_{36}
			\end{array}
			& c^{}_{34} c^{}_{35} c^{}_{36} 
		\end{pmatrix} \;.
	\end{eqnarray}
In addition, the mixing matrix ${\bf R}$ in the CC interaction of heavy Majorana neutrinos is
	\begin{eqnarray}
		{\bf R} = \begin{pmatrix}
			\hat{s}^{*}_{14} c^{}_{15} c^{}_{16} &  \hat{s}^{*}_{15} c^{}_{16} & \hat{s}^{*}_{16} \\
			\begin{array}{l}
				-\hat{s}^{*}_{14} \hat{s}^{}_{15} \hat{s}^{*}_{25} c^{}_{26} - \hat{s}^{*}_{14} c^{}_{15} \hat{s}^{}_{16} \hat{s}^{*}_{26} \\
				+ c^{}_{14} \hat{s}^{*}_{24} c^{}_{25} c^{}_{26}
			\end{array} 
			& c^{}_{15} \hat{s}^{*}_{25} c^{}_{26} - \hat{s}^{*}_{15} \hat{s}^{}_{16} \hat{s}^{*}_{26} & c^{}_{16} \hat{s}^{*}_{26} \\
			\begin{array}{l}
				- \hat{s}^{*}_{14} \hat{s}^{}_{15} c^{}_{25} \hat{s}^{*}_{35} c^{}_{36} + \hat{s}^{*}_{14} \hat{s}^{}_{15} \hat{s}^{*}_{25} \hat{s}^{}_{26} \hat{s}^{*}_{36} \\
				-\hat{s}^{*}_{14} c^{}_{15} \hat{s}^{}_{16} c^{}_{26} \hat{s}^{*}_{36} - c^{}_{14} \hat{s}^{*}_{24} \hat{s}^{}_{25} \hat{s}^{*}_{35} c^{}_{36} \\
				-c^{}_{14} \hat{s}^{*}_{24} c^{}_{25} \hat{s}^{}_{26} \hat{s}^{*}_{36} + c^{}_{14} c^{}_{24} \hat{s}^{*}_{34} c^{}_{35} c^{}_{36}
			\end{array} 
			& \begin{array}{l}
				-\hat{s}^{*}_{15} \hat{s}^{}_{16} c^{}_{26} \hat{s}^{*}_{36} - c^{}_{15} \hat{s}^{*}_{25} \hat{s}^{}_{26} \hat{s}^{*}_{36} \\ + c^{}_{15} c^{}_{25} \hat{s}^{*}_{35} c^{}_{36}
			\end{array} & c^{}_{16} c^{}_{26} \hat{s}^{*}_{36}
		\end{pmatrix} \;.
	\end{eqnarray}	
Since the matrices ${\bf D}$, ${\bf B}$ and ${\bf U}_0^{}$ are absent in all interaction terms, we shall not show their explicit expressions, which however can be found in Ref.~\cite{Xing:2011ur}. In total, there are eighteen physical parameters in the type-I seesaw model, which can be chosen as nine rotation angles $\theta_{ij}^{}$ (for $i=1,2,3$ and $j=4,5,6$) and six independent CP-violating phases $\alpha_i^{} \equiv \delta_{i4}^{} - \delta_{i5}^{}, \beta_i^{} \equiv \delta_{i5}^{} - \delta_{i6}^{}$ (for $i=1,2,3$) in ${\bf R}$ and three heavy Majorana neutrino masses $\left\{M_1^{}, M_2^{}, M_3^{}\right\}$. With the help of these {\it original} parameters~\cite{Xing:2023kdj,Xing:2024xwb,Xing:2024gmy} and the exact seesaw relation, one can calculate all other physical quantities. To simplify the formulas, we also introduce $\gamma_i^{} \equiv \delta_{i6}^{} - \delta_{i4}^{}$ (for $i=1,2,3$), which are not independent parameters but satisfy the identities $\alpha_i^{} + \beta_i^{} + \gamma_i^{} = 0$.
	
From the RGEs of ${\cal B}_{\alpha i}^{}$ (for $\alpha=e,\mu,\tau$ and $i=4,5,6$), i.e., the matrix elements of ${\bf R}$, we can extract the RGEs of physical mixing angles $\theta_{ij}^{}$ and CP-violating phases $\delta_{ij}^{}$. First, for $(i, j) = (1, 4), (1, 5), (1, 6), (2, 6), (3, 6)$, it is straightforward to obtain
	\begin{eqnarray}
		\dot{\theta}_{14}^{} &=& c_{14}^{-1} c^{-1}_{15} c^{-1}_{16} \left[{\rm Re} \dot{\cal B}_{e4}^{} \cos\delta_{14}^{} -\text{\rm Im}\dot{\cal B}_{e4}^{} \sin\delta_{14} + s_{14}^{} t_{15}^{} \left({\rm Re}\dot{\cal B}_{e5}^{} \cos\delta_{15}^{} - {\rm Im}\dot{\cal B}_{e5}^{} \sin\delta_{15}^{}\right) \right. \nonumber \\
		&& \left. + s_{14}^{} c^{-1}_{15} t_{16}^{} \left({\rm Re} \dot{\cal B}_{e6}^{} \cos\delta_{16}^{} - {\rm Im} \dot{\cal B}_{e6}^{} \sin\delta_{16}\right)\right] \;, \nonumber \\
		\dot{\theta}_{15}^{} &=& c^{-1}_{15} c^{-1}_{16} \left[ {\rm Re} \dot{\cal B}_{e5}^{} \cos\delta_{15}^{} - {\rm Im} \dot{\cal B}_{e5}^{} \sin\delta_{15} + s_{15}^{} t_{16}^{} \left({\rm Re} \dot{\cal B}_{e6}^{} \cos\delta_{16} - {\rm Im} \dot{\cal B}_{e6}^{} \sin\delta_{16}^{}\right) \right] \;, \nonumber \\
		\dot{\theta}_{16}^{} &=& c^{-1}_{16} \left({\rm Re} \dot{\cal B}_{e6}^{} \cos\delta_{16}^{} - {\rm Im} \dot{\cal B}_{e6}^{} \sin\delta_{16} \right) \;, \nonumber \\
		\dot{\theta}_{26}^{} &=& c^{-1}_{16} c^{-1}_{26} \left[t_{16}^{} s_{26}^{} \left({\rm Re} \dot{\cal B}_{e6}^{} \cos\delta_{16}^{} - {\rm Im} \dot{\cal B}_{e6}^{} \sin\delta_{16}^{} \right) +  {\rm Re} \dot{\cal B}_{\mu 6}^{} \cos\delta_{26}^{} - {\rm Im} \dot{\cal B}_{\mu 6}^{} \sin\delta_{26}^{} \right] \;, \nonumber \\
		\dot{\theta}_{36}^{} &=& c^{-1}_{16} c^{-1}_{26} c^{-1}_{36} \left[ t_{16}^{} c^{-1}_{26} s_{36}^{} \left({\rm Re} \dot{\cal B}_{e6}^{} \cos\delta_{16}^{} - {\rm Im} \dot{\cal B}_{e6}^{} \sin\delta_{16}^{} \right)   \right. \nonumber \\
		&& \left. + t_{26}^{} s_{36}^{} \left({\rm Re} \dot{\cal B}_{\mu 6}^{} \cos\delta_{26}^{} - {\rm Im} \dot{\cal B}_{\mu 6}^{} \sin\delta_{26}^{}\right) + {\rm Re} \dot{\cal B}_{\tau 6}^{} \cos\delta_{36}^{} - {\rm Im} \dot{\cal B}_{\tau 6}^{} \sin\delta_{36}^{} \right] \;,
	\end{eqnarray}
and
	\begin{eqnarray}
		\dot{\delta}_{14}^{} &=& - s^{-1}_{14} c^{-1}_{15} c^{-1}_{16} \left({\rm Re} \dot{\cal B}_{e4}^{} \sin\delta_{14}^{} + {\rm Im} \dot{\cal B}_{e4}^{} \cos\delta_{14}^{}  \right) \;, \nonumber \\
		\dot{\delta}_{15}^{} &=& - s^{-1}_{15} c^{-1}_{16} \left({\rm Re} \dot{\cal B}_{e5}^{} \sin\delta_{15}^{} + {\rm Im} \dot{\cal B}_{e5}^{} \cos\delta_{15}^{}\right) \nonumber \;,\\
		\dot{\delta}_{16}^{} &=& - s^{-1}_{16} \left({\rm Re} \dot{\cal B}_{e6}^{} \sin\delta_{16}^{} + {\rm Im} \dot{\cal B}_{e6}^{} \cos\delta_{16}^{}\right) \;, \nonumber \\
		\dot{\delta}_{26}^{} &=& - c^{-1}_{16} s^{-1}_{26} \left({\rm Re} \dot{\cal B}_{\mu 6}^{} \sin\delta_{26}^{} + {\rm Im} \dot{\cal B}_{\mu 6}^{} \cos\delta_{26}^{}\right) \;, \nonumber \\
		\dot{\delta}_{36}^{} &=& - c^{-1}_{16} c^{-1}_{26} s^{-1}_{36} \left({\rm Re} \dot{\cal B}_{\tau 6}^{} \sin\delta_{36}^{} + {\rm Im} \dot{\cal B}_{\tau 6}^{} \cos\delta_{36}^{}\right) \;,
	\end{eqnarray}
where $s_{ij}^{} \equiv \sin\theta_{ij}^{}, c_{ij}^{} \equiv \cos\theta_{ij}^{}$ and $t_{ij}^{} \equiv \tan\theta_{ij}^{}$ should be noticed. Then, with the help of the above equations, the RGEs of $\{\theta_{25}^{},\theta_{35}^{}\}$ and $\{\delta_{25}^{},\delta_{35}^{}\}$ can be found
	\begin{eqnarray}
		\dot{\theta}_{25}^{} &=& c^{-1}_{15} c^{-1}_{25} c^{-1}_{26} \left({\rm Re} \dot{\cal B}_{\mu 5}^{} \cos\delta_{25}^{} - {\rm Im} \dot{\cal B}_{\mu 5}^{} \sin\delta_{25}^{}\right) + \dot{\theta}^{}_{15} \left[s^{}_{16} c^{-1}_{25} t^{}_{26} \cos(\beta_1^{}-\beta_2^{}) + t^{}_{15} t^{}_{25} \right]  \nonumber \\
		&&  + \dot{\theta}_{16}^{} t^{}_{15} c^{}_{16} c^{-1}_{25} t^{}_{26} \cos(\beta_1^{}-\beta_2^{}) + \dot{\theta}^{}_{26} \left[t^{}_{15} s^{}_{16} c^{-1}_{25} \cos(\beta_1^{}-\beta_2^{}) + t^{}_{25} t^{}_{26} \right] \nonumber \\
		&& - t^{}_{15} s^{}_{16} c^{-1}_{25} t^{}_{26} \left(\dot{\delta}^{}_{15} - \dot{\delta}^{}_{16} + \dot{\delta}^{}_{26}\right) \sin (\beta_1^{}-\beta_2^{})  \;, \nonumber \\
		\dot{\theta}_{35}^{} &=& c^{-1}_{15} c^{-1}_{25} c^{-1}_{35} c^{-1}_{36} \left({\rm Re} \dot{\cal B}^{}_{\tau 5} \cos\delta^{}_{35} - {\rm Im} \dot{\cal B}^{}_{\tau 5} \sin\delta^{}_{35} \right) \nonumber \\
		&& + \dot{\theta}^{}_{15}  \left\{c^{-1}_{35} t^{}_{36} \left[s^{}_{16} c^{-1}_{25} c^{}_{26} \cos(\beta_1^{}-\beta_3^{}) - t^{}_{15} t^{}_{25} s^{}_{26} \cos(\beta_2^{}-\beta_3^{}) \right] + t^{}_{15} t^{}_{35} \right\} \nonumber \\
		&& + \dot{\theta}^{}_{16} t^{}_{15} c^{}_{16} c^{-1}_{25} c^{}_{26} c^{-1}_{35} t^{}_{36} \cos(\beta_1^{}-\beta_3^{}) + \dot{\theta}^{}_{25} \left[s^{}_{26} c^{-1}_{35} t^{}_{36} \cos(\beta_2^{}-\beta_3^{}) + t^{}_{25} t^{}_{35} \right] \nonumber \\
		&& + \dot{\theta}^{}_{26} c^{-1}_{25} c^{-1}_{35} t^{}_{36} \left[s^{}_{25} c^{}_{26} \cos(\beta_2^{}-\beta_3^{}) - t^{}_{15} s^{}_{16} s^{}_{26} \cos(\beta_1^{}-\beta_3^{}) \right] \nonumber \\
		&& + \dot{\theta}^{}_{36} \left\{ c^{-1}_{35} \left[t^{}_{15} s^{}_{16} c^{-1}_{25} c^{}_{26} \cos (\beta_1^{}-\beta_3^{})+t^{}_{25} s^{}_{26} \cos (\beta_2^{}-\beta_3^{}) \right] + t^{}_{35}  t^{}_{36} \right\} \nonumber \\
		&& - c^{-1}_{25} c^{-1}_{35} t^{}_{36} \left[ t^{}_{15} s^{}_{16} c^{}_{26} \left(\dot{\delta}^{}_{15}-\dot{\delta}^{}_{16}+\dot{\delta}^{}_{36}\right) \sin(\beta_1^{}-\beta_3^{}) \right. \nonumber \\
		&& \left. + s^{}_{25} s^{}_{26} \left(\dot{\delta}^{}_{25}-\dot{\delta}^{}_{26}+\dot{\delta}^{}_{36}\right) \sin(\beta_2^{}-\beta_3^{}) \right]  \;, 
	\end{eqnarray}
	and
	\begin{eqnarray}
		\dot{\delta}_{25}^{} &=&  - c^{-1}_{15} s^{-1}_{25} c^{-1}_{26} \left({\rm Re} \dot{\cal B}_{\mu 5}^{} \sin\delta_{25}^{} + {\rm Im} \dot{\cal B}_{\mu 5}^{} \cos\delta_{25}^{}\right) + \dot{\theta}_{15}^{} s^{}_{16} s^{-1}_{25} t^{}_{26} \sin(\beta_1^{}-\beta_2^{})  \nonumber\\
		&&   + \dot{\theta}_{16}^{} t^{}_{15}  c^{}_{16} s^{-1}_{25} t^{}_{26} \sin(\beta_1^{}-\beta_2^{}) + \dot{\theta}_{26}^{} t^{}_{15} s^{}_{16} s^{-1}_{25} \sin(\beta_1^{}-\beta_2^{}) \nonumber \\
		&& + t^{}_{15} s^{}_{16} s^{-1}_{25} t^{}_{26} \left(\dot{\delta}^{}_{15} - \dot{\delta}^{}_{16} + \dot{\delta}^{}_{26}\right) \cos (\beta_1^{}-\beta_2^{}) \;, \nonumber \\
		\dot{\delta}_{35}^{} &=& - c^{-1}_{15} c^{-1}_{25} s^{-1}_{35} c^{-1}_{36} \left({\rm Re} \dot{\cal B}^{}_{\tau 5} \sin\delta^{}_{35} + {\rm Im} \dot{\cal B}^{}_{\tau 5} \cos\delta^{}_{35} \right) \nonumber \\
		&& + \dot{\theta}^{}_{15} s^{-1}_{35} t^{}_{36} \left[s^{}_{16} c^{-1}_{25} c^{}_{26} \sin(\beta_1^{}-\beta_3^{}) - t^{}_{15} t^{}_{25} s^{}_{26} \sin(\beta_2^{}-\beta_3^{}) \right] \nonumber \\
		&& + \dot{\theta}^{}_{16} t^{}_{15} c^{}_{16} c^{-1}_{25} c^{}_{26} s^{-1}_{35} t^{}_{36} \sin (\beta_1^{}-\beta_3^{}) + \dot{\theta}^{}_{25} s^{}_{26} s^{-1}_{35} t^{}_{36} \sin (\beta_2^{}-\beta_3^{}) \nonumber \\
		&& + \dot{\theta}^{}_{26} s^{-1}_{35} t^{}_{36} \left[t^{}_{25} c^{}_{26} \sin (\beta_2^{}-\beta_3^{})-t^{}_{15} s^{}_{16} c^{-1}_{25} s^{}_{26} \sin (\beta_1^{}-\beta_3^{}) \right] \nonumber \\
		&& + \dot{\theta}^{}_{36} s^{-1}_{35} \left[ t^{}_{15} s^{}_{16} c^{-1}_{25} c^{}_{26} \sin (\beta_1^{}-\beta_3^{})+t^{}_{25} s^{}_{26} \sin (\beta_2^{}-\beta_3^{}) \right] \nonumber \\
		&& + c^{-1}_{25} s^{-1}_{35} t^{}_{36} \left[ t^{}_{15} s^{}_{16} c^{}_{26} \left(\dot{\delta}^{}_{15}-\dot{\delta}^{}_{16}+\dot{\delta}^{}_{36}\right) \cos (\beta_1^{}-\beta_3^{}) \right. \nonumber \\
		&& \left. +s^{}_{25} s^{}_{26} \left(\dot{\delta}^{}_{25}-\dot{\delta}^{}_{26}+\dot{\delta}^{}_{36}\right) \cos (\beta_2^{}-\beta_3^{}) \right] \;. 
	\end{eqnarray}
Furthermore, the RGEs of $\theta_{24}^{}$ and $\delta_{24}^{}$ can then be calculated
	\begin{eqnarray}
		\dot{\theta}_{24}^{} &=& c^{-1}_{14} c_{24}^{-1} c^{-1}_{25} c^{-1}_{26} \left({\rm Re} \dot{\cal B}_{\mu 4}^{} \cos\delta_{24}^{} - {\rm Im} \dot{\cal B}_{\mu 4}^{} \sin\delta_{24}^{} \right)  \nonumber \\
		&& + \dot{\theta}_{14}^{} c_{24}^{-1} \left[s^{}_{15} t^{}_{25} \cos(\alpha_1^{}-\alpha_2^{}) + c^{}_{15} s^{}_{16} c^{-1}_{25} t^{}_{26} \cos (\gamma_2^{}-\gamma_1^{}) + t^{}_{14} s^{}_{24} \right] \nonumber \\
		&& + \dot{\theta}^{}_{15} t^{}_{14} c_{24}^{-1} c^{-1}_{25} \left[c^{}_{15} s^{}_{25} \cos(\alpha_1^{}-\alpha_2^{}) - s^{}_{15} s^{}_{16} t^{}_{26} \cos(\gamma_2^{}-\gamma_1^{}) \right] \nonumber \\
		&& + \dot{\theta}^{}_{16} t^{}_{14} c^{}_{15} c^{}_{16} c^{-1}_{24} c^{-1}_{25} t^{}_{26} \cos(\gamma_2^{}-\gamma_1^{})  + \dot{\theta}^{}_{25} c_{24}^{-1} \left[t^{}_{14} s^{}_{15}  \cos(\alpha_1^{}-\alpha_2^{}) + s^{}_{24} t^{}_{25} \right] \nonumber \\
		&& + \dot{\theta}^{}_{26} c_{24}^{-1} \left[ t^{}_{14}  c^{}_{15} s^{}_{16} c^{-1}_{25} \cos(\gamma_2^{}-\gamma_1^{}) - t^{}_{14}  s^{}_{15} t^{}_{25} t^{}_{26} \cos(\alpha_1^{}-\alpha_2^{})  + s^{}_{24} t^{}_{26} \right] \nonumber \\
		&& + t^{}_{14} c_{24}^{-1} c^{-1}_{25} \left[ s^{}_{15} s^{}_{25} \left(\dot{\delta}^{}_{15} -\dot{\delta}_{14}^{} - \dot{\delta}^{}_{25}\right) \sin (\alpha_1^{}-\alpha_2^{})  \right. \nonumber \\
		&& \left. + c^{}_{15} s^{}_{16} t^{}_{26} \left(\dot{\delta}^{}_{16} -\dot{\delta}_{14}^{} - \dot{\delta}^{}_{26}\right) \sin (\gamma_2^{}-\gamma_1^{}) \right]  \;, \nonumber \\ 
		\dot{\delta}_{24}^{} &=&  - c^{-1}_{14} s_{24}^{-1} c_{25}^{-1} c^{-1}_{26} \left( {\rm Re} \dot{\cal B}_{\mu 4}^{} \sin\delta^{}_{24} + {\rm Im} \dot{\cal B}^{}_{\mu 4} \cos\delta^{}_{24} \right) \nonumber \\
		&& + \dot{\theta}^{}_{14} s_{24}^{-1} c_{25}^{-1}\left[ s^{}_{15} s^{}_{25} \sin (\alpha_1^{}-\alpha_2^{}) + c^{}_{15} s^{}_{16} t^{}_{26} \sin (\gamma_2^{}-\gamma_1^{}) \right] \nonumber \\
		&& + \dot{\theta}^{}_{15} t^{}_{14} s_{24}^{-1} c_{25}^{-1} \left[c^{}_{15} s^{}_{25} \sin (\alpha_1^{}-\alpha_2^{})-s^{}_{15} s^{}_{16} t^{}_{26} \sin (\gamma_2^{}-\gamma_1^{}) \right] \nonumber \\
		&& + \dot{\theta}^{}_{16} t^{}_{14} c^{}_{15} c^{}_{16} s_{24}^{-1} c_{25}^{-1} t^{}_{26} \sin(\gamma_2^{}-\gamma_1^{}) + \dot{\theta}^{}_{25} t^{}_{14}  s^{}_{15} s_{24}^{-1} \sin(\alpha_1^{}-\alpha_2^{})   \nonumber \\
		&& + \dot{\theta}^{}_{26} t^{}_{14}s_{24}^{-1} c_{25}^{-1} \left[c^{}_{15} s^{}_{16} \sin (\gamma_2^{}-\gamma_1^{})-s^{}_{15} s^{}_{25} t^{}_{26} \sin (\alpha_1^{}-\alpha_2^{}) \right] \nonumber \\
		&& + t^{}_{14} s_{24}^{-1} c_{25}^{-1} \left[ s^{}_{15} s^{}_{25} \left(\dot{\delta}^{}_{25} - \dot{\delta}^{}_{15} + \dot{\delta}^{}_{14} \right) \cos(\alpha_1^{}-\alpha_2^{}) \right. \nonumber \\
		&& \left. +c^{}_{15} s^{}_{16} t^{}_{26} \left(\dot{\delta}^{}_{26} - \dot{\delta}^{}_{16} + \dot{\delta}^{}_{14} \right) \cos(\gamma_2^{}-\gamma_1^{}) \right]  \;.
	\end{eqnarray}
Finally, after some straightforward but tedious calculations, we get
	\begin{eqnarray}
		\dot{\theta}_{34}^{} &=& c^{-1}_{14} c^{-1}_{24} c^{-1}_{34} c^{-1}_{35} c^{-1}_{36} \left({\rm Re} \dot{\cal B}^{}_{\tau 4} \cos\delta^{}_{34} - {\rm Im} \dot{\cal B}^{}_{\tau 4} \sin\delta^{}_{34} \right)  \nonumber \\
		&& + \dot{\theta}^{}_{14} \left\{ t^{}_{14} t^{}_{34} + c^{-1}_{24} c^{-1}_{34} \left\{ s^{}_{15} c^{}_{25}  t^{}_{35} \cos(\alpha_1^{}-\alpha_3^{})  \right.\right.\nonumber \\
		&& \left. + c^{-1}_{35} t^{}_{36} [c^{}_{15} s^{}_{16} c^{}_{26} \cos (\gamma_3^{}-\gamma_1^{}) - s^{}_{15} s^{}_{25} s^{}_{26} \cos (\alpha_1^{}+\beta_2^{}+\gamma_3^{}) ] \right\}  \nonumber \\
		&& \left. - t^{}_{14} t^{}_{24} c^{-1}_{34} \left[s^{}_{25} t^{}_{35} \cos(\alpha_2^{}-\alpha_3^{}) + c^{}_{25} s^{}_{26} c^{-1}_{35}  t^{}_{36}\cos (\gamma_3^{}-\gamma_2^{}) \right] \right\}  \nonumber \\
		&& + \dot{\theta}^{}_{15} t^{}_{14} c^{-1}_{24} c^{-1}_{34} c^{-1}_{35} \left[c^{}_{15} c^{}_{25} s^{}_{35} \cos(\alpha_1^{}-\alpha_3^{}) - s^{}_{15} s^{}_{16} c^{}_{26} t^{}_{36} \cos(\gamma_3^{}-\gamma_1^{})  \right. \nonumber \\
		&& \left. - c^{}_{15} s^{}_{25} s^{}_{26} t^{}_{36} \cos(\alpha_1^{}+\beta_2^{}+\gamma_3^{}) \right] + \dot{\theta}^{}_{16} t^{}_{14} c^{}_{15} c^{}_{16} c^{-1}_{24} c^{}_{26} c^{-1}_{34} c^{-1}_{35}  t^{}_{36} \cos(\gamma_3^{}-\gamma_1^{}) \nonumber \\
		&& + \dot{\theta}^{}_{24} c^{-1}_{34} \left[ t^{}_{24} s^{}_{34} + s^{}_{25} t^{}_{35} \cos(\delta^{}_{24}-\delta^{}_{25} - \delta_{34}^{}+\delta^{}_{35}) + c^{}_{25} s^{}_{26} c^{-1}_{35} t^{}_{36} \cos(\gamma_3^{}-\gamma_2^{}) \right] \nonumber \\
		&& + \dot{\theta}^{}_{25} c^{-1}_{24} c^{-1}_{34} c^{-1}_{35} \left\{ s^{}_{24} c^{}_{25} s^{}_{35} \cos(\alpha_2^{}-\alpha_3^{}) - t^{}_{14} s^{}_{15} s^{}_{25} s^{}_{35} \cos(\alpha_1^{}-\alpha_3^{})  \right.\nonumber \\
		&& \left. -s^{}_{26} t^{}_{36} \left[s^{}_{24} s^{}_{25} \cos(\gamma_3^{}-\gamma_2^{}) + t^{}_{14} s^{}_{15} c^{}_{25} \cos(\alpha_1^{}+\beta_2^{}+\gamma_3^{}) \right] \right\}  \nonumber \\
		&& + \dot{\theta}^{}_{26} c^{-1}_{24} c^{-1}_{34} c^{-1}_{35} t^{}_{36} \left[ s^{}_{24} c^{}_{25} c^{}_{26} \cos(\gamma_3^{}-\gamma_2^{}) \right. \nonumber\\
		&& \left. -t^{}_{14} s^{}_{15} s^{}_{25} c^{}_{26} \cos(\alpha_1^{}+\beta_2^{}+\gamma_3^{}) - t^{}_{14} c^{}_{15} s^{}_{16} s^{}_{26} \cos(\gamma_3^{}-\gamma_1^{}) \right]  \nonumber \\
		&& + \dot{\theta}^{}_{35} c^{-1}_{34} \left[t^{}_{14} s^{}_{15} c^{-1}_{24} c^{}_{25} \cos(\alpha_1^{}-\alpha_3^{}) + t^{}_{24} s^{}_{25} \cos(\alpha_2^{}-\alpha_3^{})  + s^{}_{34} t^{}_{35} \right]  \nonumber \\
		&& + \dot{\theta}^{}_{36} c^{-1}_{34} \left[ t^{}_{14} c^{}_{15} s^{}_{16} c^{-1}_{24} c^{}_{26} c^{-1}_{35} \cos(\gamma_3^{}-\gamma_1^{}) - t^{}_{14} s^{}_{15} c^{-1}_{24} s^{}_{25} s^{}_{26}  c^{-1}_{35} \cos(\alpha_1^{}+\beta_2^{}+\gamma_3^{}) \right. \nonumber \\
		&& - t^{}_{14} s^{}_{15} c^{-1}_{24} c^{}_{25} t^{}_{35} t^{}_{36} \cos(\alpha_1^{}-\alpha_3^{}) + t^{}_{24} c^{}_{25} s^{}_{26} c^{-1}_{35} \cos(\gamma_3^{}-\gamma_2^{}) \nonumber \\
		&& \left. - t^{}_{24} s^{}_{25} t^{}_{35} t^{}_{36} \cos(\alpha_2^{}-\alpha_3^{}) + s_{34}^{} t^{}_{36} \right] \nonumber \\
		&& - \dot{\delta}^{}_{14} t^{}_{14}  c^{-1}_{24} c^{-1}_{34} \left\{s^{}_{15} c^{}_{25} t^{}_{35} \sin(\alpha_1^{}-\alpha_3^{}) + c^{-1}_{35} t^{}_{36} \left[c^{}_{15} s^{}_{16} c^{}_{26} \sin (\gamma_3^{}-\gamma_1^{}) \right.\right.\nonumber \\
		&& \left.\left. - s^{}_{15} s^{}_{25} s^{}_{26} \sin(\alpha_1^{}+\beta_2^{}+\gamma_3^{}) \right] \right\}  \nonumber \\
		&& + \dot{\delta}^{}_{15} t^{}_{14} s^{}_{15} c^{-1}_{24} c^{-1}_{34} c^{-1}_{35} \left[ c^{}_{25} s^{}_{35} \sin(\alpha_1^{}-\alpha_3^{}) - s^{}_{25} s^{}_{26} t^{}_{36} \sin(\alpha_1^{}+\beta_2^{}+\gamma_3^{}) \right]  \nonumber \\
		&& + \dot{\delta}^{}_{16} t^{}_{14} c^{}_{15} s^{}_{16} c^{-1}_{24} c^{}_{26}  c^{-1}_{34} c^{-1}_{35} t^{}_{36} \sin(\gamma_3^{}-\gamma_1^{})     \nonumber \\
		&& - \dot{\delta}^{}_{24} t^{}_{24} c^{-1}_{34} c^{-1}_{35} \left[s^{}_{25} s^{}_{35} \sin(\alpha_2^{}-\alpha_3^{}) + c^{}_{25} s^{}_{26} t^{}_{36} \sin(\gamma_3^{}-\gamma_2^{}) \right]  \nonumber \\
		&& + \dot{\delta}^{}_{25} s^{}_{25} c^{-1}_{34} c^{-1}_{35} \left[t^{}_{24} s^{}_{35} \sin(\alpha_2^{}-\alpha_3^{}) + t^{}_{14} s^{}_{15} c^{-1}_{24} s^{}_{26} t^{}_{36} \sin(\alpha_1^{}+\beta_2^{}+\gamma_3^{}) \right]  \nonumber \\
		&& + \dot{\delta}^{}_{26} s^{}_{26} c^{-1}_{34} c^{-1}_{35} t^{}_{36} \left[t^{}_{24} c^{}_{25} \sin(\gamma_3^{}-\gamma_2^{}) - t^{}_{14} s^{}_{15} c^{-1}_{24} s^{}_{25}\sin (\alpha_1^{}+\beta_2^{}+\gamma_3^{}) \right]   \nonumber \\
		&& - \dot{\delta}^{}_{35} c^{-1}_{24} c^{-1}_{34} t^{}_{35} \left[s^{}_{24} s^{}_{25} \sin(\alpha_2^{}-\alpha_3^{}) + t^{}_{14} s^{}_{15} c^{}_{25} \sin(\alpha_1^{}-\alpha_3^{}) \right] \nonumber \\
		&& - \dot{\delta}^{}_{36} c^{-1}_{24} c^{-1}_{34} c^{-1}_{35} t^{}_{36} \left[  s^{}_{24} c^{}_{25} s^{}_{26} \sin(\gamma_3^{}-\gamma_2^{}) \right.\nonumber\\
		&& \left. + t^{}_{14} c^{}_{15} s^{}_{16} c^{}_{26} \sin(\gamma_3^{}-\gamma_1^{}) - t^{}_{14} s^{}_{15} s^{}_{25} s^{}_{26} \sin(\alpha_1^{}+\beta_2^{}+\gamma_3^{}) \right]   \;, 
	\end{eqnarray}
	and
	\begin{eqnarray}
		\dot{\delta}_{34}^{} &=& - c^{-1}_{14} c^{-1}_{24} s^{-1}_{34} c^{-1}_{35} c^{-1}_{36} \left({\rm Re}\dot{\cal B}^{}_{\tau 4} \sin\delta^{}_{34} + {\rm Im} \dot{\cal B}^{}_{\tau 4} \cos\delta^{}_{34} \right) \nonumber \\
		&& + \dot{\theta}^{}_{14} s^{-1}_{34} \left\{ s^{}_{15} c^{-1}_{24} c^{}_{25} t^{}_{35} \sin(\alpha_1^{}-\alpha_3^{}) - t^{}_{14} t^{}_{24} s^{}_{25} t^{}_{35} \sin(\alpha_2^{}-\alpha_3^{})  \right. \nonumber \\
		&& + c^{-1}_{24} c^{-1}_{35} t^{}_{36} \left[c^{}_{15} s^{}_{16} c^{}_{26}  \sin (\gamma_3^{}-\gamma_1^{})  \right. \nonumber \\
		&& \left.\left. - s^{}_{15} s^{}_{25} s^{}_{26} \sin(\alpha_1^{}+\beta_2^{}+\gamma_3^{}) - t^{}_{14} s^{}_{24} c^{}_{25} s^{}_{26} \sin(\gamma_3^{}-\gamma_2^{}) \right] \right\}  \nonumber \\
		&& + \dot{\theta}^{}_{15} t^{}_{14} c^{-1}_{24} s^{-1}_{34} c^{-1}_{35} \left[ c^{}_{15} c^{}_{25} s^{}_{35} \sin(\alpha_1^{}-\alpha_3^{}) \right. \nonumber \\
		&& \left. - s^{}_{15} s^{}_{16} c^{}_{26} t^{}_{36} \sin(\gamma_3^{}-\gamma_1^{}) - c^{}_{15} s^{}_{25} s^{}_{26} t^{}_{36} \sin(\alpha_1^{}+\beta_2^{}+\gamma_3^{})  \right]  \nonumber \\
		&& + \dot{\theta}^{}_{16} t^{}_{14} c^{}_{15} c^{}_{16} c^{-1}_{24} c^{}_{26} s^{-1}_{34} c^{-1}_{35} t^{}_{36} \sin(\gamma_3^{}-\gamma_1^{})    \nonumber \\
		&& + \dot{\theta}^{}_{24} s^{-1}_{34} \left[s^{}_{25} t^{}_{35} \sin(\alpha_2^{}-\alpha_3^{}) + c^{}_{25} s^{}_{26} c^{-1}_{35} t^{}_{36} \sin (\gamma_3^{}-\gamma_2^{})  \right]  \nonumber \\
		&& - \dot{\theta}^{}_{25} s^{-1}_{34} c^{-1}_{35} \left\{ t^{}_{24} s^{}_{25} s^{}_{26}  t^{}_{36} \sin(\gamma_3^{}-\gamma_2^{}) - t^{}_{24} c^{}_{25} s^{}_{35} \sin(\alpha_2^{}-\alpha_3^{}) \right.\nonumber \\
		&& \left. + t^{}_{14} s^{}_{15} c^{-1}_{24} \left[s^{}_{25} s^{}_{35} \sin(\alpha_1^{}-\alpha_3^{}) + c^{}_{25} s^{}_{26} t^{}_{36} \sin(\alpha_1^{}+\beta_2^{}+\gamma_3^{})  \right]  \right\}  \nonumber \\
		&& + \dot{\theta}^{}_{26} c^{-1}_{24} s^{-1}_{34} c^{-1}_{35} t^{}_{36} \left[s^{}_{24} c^{}_{25} c^{}_{26} \sin(\gamma_3^{}-\gamma_2^{})  \right. \nonumber \\
		&& \left. -  t^{}_{14} s^{}_{15} s^{}_{25} c^{}_{26} \sin(\alpha_1^{}+\beta_2^{}+\gamma_3^{}) - t^{}_{14} c^{}_{15} s^{}_{16} s^{}_{26} \sin(\gamma_3^{}-\gamma_1^{}) \right]   \nonumber \\
		&& + \dot{\theta}^{}_{35} s^{-1}_{34} \left[t^{}_{14} s^{}_{15} c^{-1}_{24} c^{}_{25}  \sin(\alpha_1^{}-\alpha_3^{}) + t^{}_{24} s^{}_{25} \sin(\alpha_2^{}-\alpha_3^{}) \right]  \nonumber \\
		&& + \dot{\theta}^{}_{36} s^{-1}_{34} c^{-1}_{35} \left\{ t^{}_{14} c^{}_{15} s^{}_{16} c^{-1}_{24} c^{}_{26} \sin(\gamma_3^{}-\gamma_1^{}) \right. \nonumber \\
		&&  - t^{}_{14} s^{}_{15} c^{-1}_{24} \left[s^{}_{25} s^{}_{26} \sin(\alpha_1^{}+\beta_2^{}+\gamma_3^{}) + c^{}_{25} s^{}_{35} t^{}_{36} \sin(\alpha_1^{}-\alpha_3^{}) \right]  \nonumber \\
		&& \left. + t^{}_{24} c^{}_{25} s^{}_{26}  \sin(\gamma_3^{}-\gamma_2^{}) - t^{}_{24} s^{}_{25} s^{}_{35} t^{}_{36} \sin(\alpha_2^{}-\alpha_3^{}) \right\}  \nonumber\\
		&& + \dot{\delta}^{}_{14} t^{}_{14} c^{-1}_{24} s^{-1}_{34} \left\{ s^{}_{15} c^{}_{25} t^{}_{35} \cos (\alpha_1^{}-\alpha_3^{}) \right. \nonumber \\
		&& \left. + c^{-1}_{35} t^{}_{36} \left[c^{}_{15} s^{}_{16} c^{}_{26} \cos(\gamma_3^{}-\gamma_1^{}) - s^{}_{15} s^{}_{25} s^{}_{26} \cos(\alpha_1^{}+\beta_2^{}+\gamma_3^{})  \right]  \right\}  \nonumber \\
		&& + \dot{\delta}^{}_{15} t^{}_{14} s^{}_{15} c^{-1}_{24} s^{-1}_{34} \left[s^{}_{25} s^{}_{26} c^{-1}_{35} t^{}_{36} \cos(\alpha_1^{}+\beta_2^{}+\gamma_3^{}) - c^{}_{25} t^{}_{35} \cos(\alpha_1^{}-\alpha_3^{}) \right]  \nonumber \\
		&& - \dot{\delta}^{}_{16} t^{}_{14} c^{}_{15} s^{}_{16} c^{-1}_{24} c^{}_{26} s^{-1}_{34}  c^{-1}_{35}   t^{}_{36} \cos(\gamma_3^{}-\gamma_1^{})   \nonumber \\
		&& + \dot{\delta}^{}_{24} t^{}_{24} s^{-1}_{34} \left[s^{}_{25} t^{}_{35} \cos(\alpha_2^{}-\alpha_3^{}) + c^{}_{25} s^{}_{26} c^{-1}_{35} t^{}_{36} \cos(\gamma_3^{}-\gamma_2^{}) \right]  \nonumber \\
		&& - \dot{\delta}^{}_{25} s^{}_{25} s^{-1}_{34}  \left[t^{}_{24} t^{}_{35} \cos(\alpha_2^{}-\alpha_3^{}) + t^{}_{14} s^{}_{15} c^{-1}_{24} s^{}_{26} c^{-1}_{35} t^{}_{36} \cos(\alpha_1^{}+\beta_2^{}+\gamma_3^{}) \right]  \nonumber \\
		&& + \dot{\delta}^{}_{26} s^{}_{26} s^{-1}_{34} c^{-1}_{35} t^{}_{36} \left[t^{}_{14} s^{}_{15} c^{-1}_{24} s^{}_{25} \cos(\alpha_1^{}+\beta_2^{}+\gamma_3^{}) - t^{}_{24} c^{}_{25} \cos(\gamma_3^{}-\gamma_2^{}) \right]   \nonumber \\
		&& + \dot{\delta}^{}_{35} s^{-1}_{34} t^{}_{35} \left[t^{}_{14} s^{}_{15} c^{-1}_{24} c^{}_{25} \cos(\alpha_1^{}-\alpha_3^{}) + t^{}_{24} s^{}_{25} \cos(\alpha_2^{}-\alpha_3^{}) \right]  \nonumber \\
		&& + \dot{\delta}^{}_{36} c^{-1}_{24} s^{-1}_{34} c^{-1}_{35} t^{}_{36} \left[ s^{}_{24} c^{}_{25} s^{}_{26} \cos (\gamma_3^{}-\gamma_2^{}) \right.\nonumber\\
		&& \left. + t^{}_{14} c^{}_{15} s^{}_{16} c^{}_{26} \cos(\gamma_3^{}-\gamma_1^{}) - t^{}_{14} s^{}_{15} s^{}_{25} s^{}_{26} \cos(\alpha_1^{}+\beta_2^{}+\gamma_3^{})  \right]  \;.
	\end{eqnarray}

Now that the RGEs of nine mixing angles and six CP-violating phases are complete, some comments on the results and those of neutrino masses are helpful. First, we emphasize that, although $\delta_{ij}^{}$ may appear alone or through combinations in $\gamma_i^{}$, the beta functions of physical mixing angles and CP-violating phases $\left\{\alpha_i^{},\beta_i^{}\right\}$ can all be expressed in terms of themselves after explicitly writing out the RGEs of the mixing matrix $\dot{\cal B}_{\alpha i}^{}$ and light neutrino masses $\left\{m_1^{},m_2^{},m_3^{}\right\}$ with {\it original} parameters. However, after doing so, we obtain quite lengthy expressions. In order to render the results more compact and readable, we have retained some implicit formulas, which can be further expanded out. Second, one may follow the strategy in Ref.~\cite{Xing:2024gmy} and make series expansions of ${\bf A}$ and ${\bf R}$ with respect to $s_{ij}^{}$, which are regarded as small parameters due to current experimental constraints. It is worthwhile to stress that since the {\it derivative} parameters, such as $\left\{m_1^{},m_2^{},m_3^{}\right\}$, appear simultaneously in the numerator and denominator in Eq.~(\ref{eq:RG_B}), one should consider all terms on the same order in such expansions.
	
\section{Summary}
\label{sec:sum}
	
In this paper, we perform a complete one-loop renormalization of the type-I seesaw model in the $\overline{\rm MS}$ scheme, and calculate for the first time the full set of RGEs for all physical parameters in the full theory. Starting with the Lagrangian in the mass basis, we derive the corresponding Feynman rules in the neutrino sector. Then, we choose the fine-structure constant $\alpha(\mu)$, physical masses $\left\{m_W^{}(\mu),m_Z^{}(\mu),m_h^{}(\mu),m_f^{}(\mu)\right\}$ and mixing matrices $\left\{{\bf V}_{}^{\rm CKM}(\mu),{\bf V}(\mu), {\bf R}(\mu)\right\}$ as input parameters. The counterterms, which contain only the UV-divergent terms in the $\overline{\rm MS}$ scheme, are fixed by imposing the renormalization conditions to obtain the UV-finite corrections. For the fermion part of our interest, the two-point self-energies are apparently modified when compared to those in the SM, so we explicitly calculate them in the $R_\xi^{}$ gauge with tadpole contributions, and derive the wave-function counterterms. Furthermore, the gauge-independent counterterms of masses and the leptonic mixing matrix are also given, which are later used to derive their RGEs. In the Euler-like parametrization of the leptonic mixing matrix, we derive the RGEs of mixing angles between light and heavy neutrinos and the CP-violating phases in ${\bf R}$, which are regarded as the {\it original} parameters of the seesaw model. The RGEs of the fine-structure constant, gauge-boson masses, the Higgs-boson mass, quark masses and CKM matrix elements are also derived, by including the contributions from massive neutrinos whenever there are, for completeness. 
	
The main result of the present paper is a theoretical framework to investigate the low-energy phenomenology of the seesaw model at the one-loop level. Such a framework will be indispensable in the forthcoming era of precision measurements of neutrino oscillation parameters, leptonic CP violation and lepton flavor or number violation. In comparison with the one-loop renormalization of the leptonic mixing matrix in Ref.~\cite{Pilaftsis:2002nc}, the RGEs of neutrino masses are also given and the modifications of the RGEs of the SM parameters due to the presence of massive Majorana neutrinos are found. Therefore, the complete set of RGEs for all physical parameters are derived and can be numerically solved. To this end, we should also perform the one-loop renormalization of the type-I seesaw mode in the on-shell scheme and establish the matching relations of on-shell parameters to physical parameters in $\overline{\rm MS}$ scheme. This is because the on-shell scheme has been widely adopted in the renormalization of the electroweak theory and the on-shell parameters can be directly extracted from experimental measurements. Just as in the case of the SM~\cite{Alam:2022cdv}, the complete on-shell renormalization of the type-I seesaw model and the comparison between different renormalization schemes are certainly of great significance and highly nontrivial. These issues will be left for the subsequent works.

	\section*{Acknowledgements}
	
	This work was supported in part by the National Natural Science Foundation of China under grant No.~12475113, by the CAS Project for Young Scientists in Basic Research (YSBR-099), and by the Scientific and Technological Innovation Program of IHEP under grant No.~E55457U2. All Feynman diagrams in this work are generated by
	{\tt FeynArts}~\cite{Hahn:2000kx}, and loop integrals are calculated with the help of {\tt Package-X}~\cite{Patel:2015tea,Patel:2016fam}.
	
	\appendix

	\section{One-loop self-energy corrections}
	
	\label{app:1-loop}
	
	To fix the wave-function and mass counterterms for charged leptons and Majorana neutrinos, one has to first calculate their 1PI two-point functions. In this Appendix, we list the analytical results for completeness. Our calculations are carried out in the general $R_\xi^{}$ gauge so that the gauge-independence can be shown clearly as a double-check of the correctness of our results. Tadpole diagrams are included, and compared with previous results in the literature. Not only the gauge-boson contributions but also those from the Higgs boson and fermions are considered in the tadpole loops.
		
	\subsection{Tadpole}
	
	\begin{figure}[t]
		\centering
		\includegraphics[scale=1]{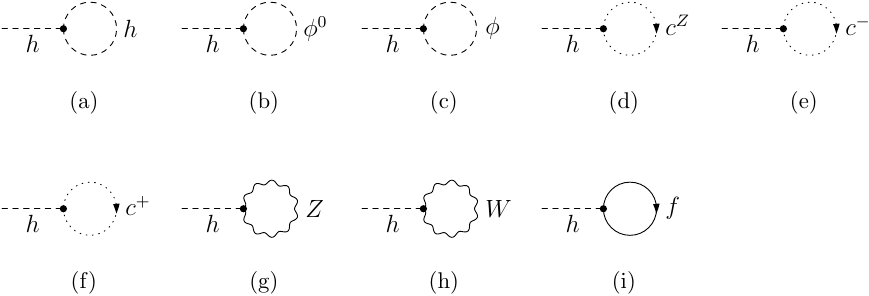}
%		\vspace{-0.7cm}
		\caption{The tadpole diagrams contributing to the two-point self-energies, where the Higgs boson $h$, the Goldstone bosons $\left\{\phi^\pm,\phi^0\right\}$, the Faddeev-Popov ghosts $\left\{c^\pm,c^Z\right\}$, the gauge bosons $\left\{W,Z\right\}$, and all the massive fermions $f$ are running in the loop.}
		\label{fig:tadpole}
	\end{figure}
	
	All tadpole diagrams are plotted in Fig.~\ref{fig:tadpole}. The total contribution is
	\begin{eqnarray}
		T &=& \frac{g m_W^{}}{2 (4\pi)^2} \left\{\vphantom{\sum_f} \frac{m_h^2}{m_W^2} \left[A^{}_0\left(\sqrt{\xi_W^{}} m_W^{}\right) + \frac{1}{2} A^{}_0\left(\sqrt{\xi_Z^{}} m_Z^{}\right) + \frac{3}{2} A^{}_0\left(m_h^{}\right)\right] + 2 (d-1) A^{}_0\left(m_W^{}\right)  \right. \nonumber \\
		&& \left. + (d-1) \frac{m_Z^2}{m_W^2} A^{}_0\left(m_Z^{}\right) - 4 \sum_{f=q,l} \frac{m_f^2}{m_W^2} A^{}_0\left(m_f^{}\right) - 4 \sum_i {\cal C}_{ii}^{} \frac{\widehat{m}_i^2}{m_W^2} A^{}_0\left(\widehat{m}_i^{}\right)  \right\} \;.
	\end{eqnarray}
	Here we adopt the PV function $A_0^{}$ to express the loop integral~\cite{Passarino:1978jh}. Note that the color factor of three for quarks should be included.
	
	\subsection{Self-energies of Charged Leptons}
	
	\begin{figure}[t]
		\centering
		\includegraphics[scale=1]{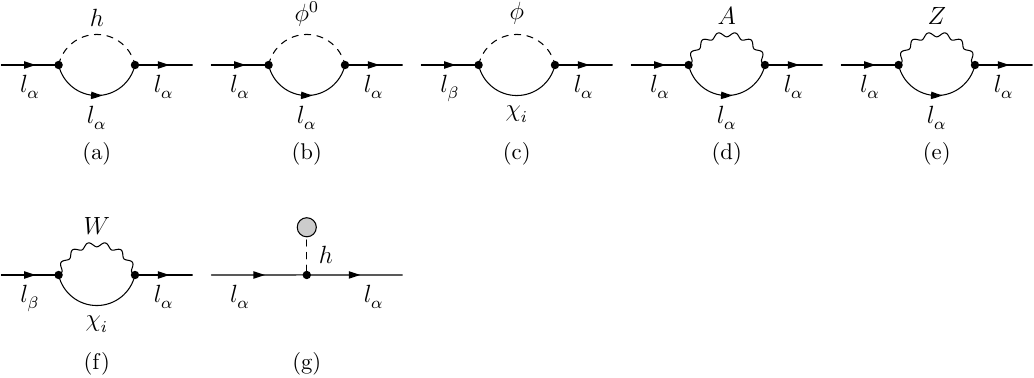}
		\vspace{-0.7cm}
		\caption{The self-energy corrections $\Sigma^{l}_{\alpha\beta} (p^2)$ to charged leptons. With the SM interactions, $\alpha=\beta$ is required in (a), (b), (d), (e) and (g), while for (c) and (f) this condition is not necessary. The tadpole contributions are summed in (g).}
		\label{fig:charged}
	\end{figure}
	
	The one-loop corrections to the self-energies of charged leptons are depicted in Fig.~\ref{fig:charged}. The left-handed, right-handed and scalar parts of the self-energy are listed below:	
	\begin{eqnarray}
		\Sigma_{\alpha \beta}^{l,{\rm L}} (p^2) &=& - \frac{g^2}{4 (4\pi)^2 } \left\{ \sum_i {\cal B}_{\alpha i}^{} {\cal B}_{\beta i}^{*} \left[\frac{p^2 - \widehat{m}_i^2}{p^2 m_W^2} \left(p^2 -\xi_W^{} m_W^2 + m_W^2\right) B_0^{}\left(p^2;m_W^{},\sqrt{\xi_W^{}} m_W^{} \right)  \right. \right. \nonumber \\
		&& + \frac{2 \left(p^2 - \widehat{m}_i^2 \right)}{m_W^2} B_1^{}\left(p^2;m_W^{},\sqrt{\xi^{}_W} m_W^{}\right) + 2 (d-2) B_1^{}\left(p^2;\widehat{m}_i^{},m_W^{}\right) \nonumber \\	
		&& + \frac{1}{p^2 m_W^2} \left(p^4 - 2 \widehat{m}_i^2 p^2 - \xi_W^{} m_W^2 p^2 + \widehat{m}_i^4 - \widehat{m}_i^2 \xi_W^{} m_W^2 \right) B_0^{}\left(p^2;\widehat{m}_i^{},\sqrt{\xi_W^{}} m_W^{} \right) \nonumber \\
		&& - \frac{1}{p^2 m_W^2} \left(p^4 - 2 \widehat{m}_i^2 p^2 - m_W^2 p^2 + \widehat{m}_i^4 - \widehat{m}_i^2 m_W^2\right) B_0^{}\left(p^2;\widehat{m}_i^{},m_W^{}\right) \nonumber \\
		&& \left. + \frac{2\widehat{m}_i^2}{m_W^2} B_1^{}\left(p^2;\widehat{m}_i^{},\sqrt{\xi^{}_W} m_W^{}\right) \right] \nonumber \\
		&& + 2 c_{\rm L}^2 \delta_{\alpha\beta}^{} \left[\frac{p^2 - m_\alpha^2}{p^2 m_W^2} \left(p^2 - \xi_Z^{} m_Z^2 + m_Z^2 \right) B_0^{}\left(p^2;m_Z^{},\sqrt{\xi_Z^{}} m_Z^{} \right) \right. \nonumber\\
		&& + \frac{2 \left(p^2 - m_\alpha^2\right)}{m_W^2} B_1^{}\left(p^2;m_Z^{},\sqrt{\xi_Z^{}} m_Z^{}\right) + \frac{2(d-2)}{c^2} B_1^{}\left(p^2;m_\alpha^{},m_Z^{}\right)  \nonumber \\
		&& + \frac{1}{p^2 m_W^2} \left(p^4 - 2 m_\alpha^2 p^2 - \xi_Z^{} m_Z^2 p^2 + m_\alpha^4 - m_\alpha^2 \xi_Z^{} m_Z^2  \right) B_0^{}\left(p^2;m_\alpha^{},\sqrt{\xi_Z^{}} m_Z^{}\right) \nonumber \\
		&& - \frac{1}{p^2 m_W^2} \left(p^4 - 2 m_\alpha^2 p^2 - m_Z^2 p^2 + m_\alpha^4 - m_\alpha^2 m_Z^2 \right) B_0^{}\left(p^2;m_\alpha^{},m_Z^{}\right) \nonumber \\
		&& \left. + \frac{m_\alpha^2}{2 c_{\rm L}^2 m_W^2} B_1^{}\left(p^2;m_\alpha^{},\sqrt{\xi_Z^{}} m_Z^{}\right) \right] \nonumber \\
		&& + 2 s^2 \delta_{\alpha\beta}^{} \left[ \frac{p^2 - m_\alpha^2}{p^2 \lambda^2} \left(p^2 -\xi_A^{} \lambda^2 + \lambda^2 \right) B_0^{}\left(p^2;\lambda,\sqrt{\xi_A^{}} \lambda \right)  \right. \nonumber \\
		&& + \frac{2\left(p^2 - m_\alpha^2\right)}{\lambda^2} B_1^{}\left(p^2;\lambda ,\sqrt{\xi_A^{}} \lambda\right) + 2 (d-2) B_1^{}\left(p^2;m_\alpha^{},\lambda\right) \nonumber \\
		&& + \frac{1}{p^2 \lambda^2} \left(p^4 - 2 m_\alpha^2 p^2 - \xi_A^{} \lambda^2 p^2 +m_\alpha^4 - m_\alpha^2 \xi_A^{} \lambda^2 \right) B_0^{}\left(p^2;m_\alpha^{},\sqrt{\xi_A^{}}\lambda \right) \nonumber \\
		&& \left. \left. - \frac{1}{p^2 \lambda^2} \left(p^4 - 2 m_\alpha^2 p^2 - \lambda^2 p^2 + m_\alpha^4 - \lambda^2 m_\alpha^2 \right) B_0^{}\left(p^2;m_\alpha^{},\lambda \right)  \right] + \frac{m_\alpha^2}{m_W^2} \delta_{\alpha\beta}^{} B_1^{}\left(p^2;m_\alpha^{},m_h^{}\right) \vphantom{\sum_i} \right\} \;, \nonumber \\
		\Sigma_{\alpha \beta}^{l,{\rm R}} &=& - \frac{g^2}{4 (4\pi)^2} \left\{ 2 \sum_i {\cal B}_{\alpha i}^{} {\cal B}_{\beta i}^{*} \frac{m_\alpha^{} m_\beta^{}}{m_W^2} B_1^{}\left(p^2;\widehat{m}_i^{},\sqrt{\xi_W^{}} m_W^{} \right) \right. \nonumber \\
		&& + 2 c_{\rm R}^2 \delta_{\alpha\beta}^{} \left[\frac{p^2 - m_\alpha^2}{p^2 m_W^2} \left(p^2 - \xi_Z^{} m_Z^2 + m_Z^2\right) B_0^{}\left(p^2;m_Z^{},\sqrt{\xi_Z^{}} m_Z^{} \right) \right. \nonumber \\
		&& + \frac{2\left(p^2 - m_\alpha^2\right)}{m_W^2}
		B_1^{}\left(p^2;m_Z^{},\sqrt{\xi_Z^{}} m_Z^{}\right) + \frac{2(d-2)}{c^2} B_1^{}\left(p^2;m_\alpha^{},m_Z^{}\right) \nonumber \\
		&& + \frac{1}{p^2 m_W^2} \left(p^4 - 2 m_\alpha^2 p^2 - \xi_Z^{} m_Z^2 p^2 + m_\alpha^4 - m_\alpha^2 \xi_Z^{} m_Z^2   \right) B_0^{}\left(p^2;m_\alpha^{},\sqrt{\xi_Z^{}} m_Z^{}\right) \nonumber \\
		&& - \frac{1}{p^2 m_W^2} \left(p^4 - 2 m_\alpha^2 p^2 - m_Z^2 p^2 + m_\alpha^4 - m_\alpha^2 m_Z^2\right) B_0^{}\left(p^2;m_\alpha^{},m_Z^{}\right) \nonumber \\
		&& \left. + \frac{m_\alpha^2}{2 c_{\rm R}^2 m_W^2} B_1^{}\left(p^2;m_\alpha^{},\sqrt{\xi_Z^{}}m_Z^{}\right) \right] \nonumber \\
		&& + 2 s^2 \delta_{\alpha\beta}^{} \left[ \frac{p^2 - m_\alpha^2}{p^2 \lambda^2} \left(p^2 - \xi_A^{} \lambda^2 + \lambda^2 \right) B_0^{}\left(p^2;\lambda,\sqrt{\xi_A^{}} \lambda\right) \right. \nonumber \\
		&& + \frac{2 \left(p^2 - m_\alpha^2\right)}{\lambda^2} B_1^{}\left(p^2;\lambda,\sqrt{\xi_A^{}} \lambda \right) + 2(d-2) B_1^{}\left(p^2;m_\alpha^{},\lambda\right) \nonumber \\
		&& + \frac{1}{p^2\lambda^2} \left(p^4 - 2 m_\alpha^2 p^2 - \xi_A^{} \lambda^2 p^2 + m_\alpha^4 - m_\alpha^2 \xi_A^{} \lambda^2 \right) B_0^{}\left(p^2;m_\alpha^{},\sqrt{\xi_A^{}} \lambda \right) \nonumber \\
		&& \left.\left. - \frac{1}{p^2\lambda^2} \left(p^4 - 2 m_\alpha^2 p^2 - \lambda^2 p^2 + m_\alpha^4 - m_\alpha^2 \lambda^2 \right) B_0^{}\left(p^2;m_\alpha^{},\lambda \right) \right] + \frac{m_\alpha^2}{m_W^2} \delta_{\alpha\beta}^{} B_1^{}\left(p^2;m_\alpha^{},m_h^{}\right) \vphantom{\sum_i} \right\} \;, \nonumber \\
		\Sigma_{\alpha \beta}^{l,{\rm D}} &=& - \frac{g^2 m_\alpha^{}}{4 (4\pi)^2} \left\{2 \sum_i {\cal B}_{\alpha i}^{} {\cal B}_{\beta i}^{*} \frac{\widehat{m}_i^2 }{m_W^2} B_0^{}\left(p^2;\widehat{m}_i^{},\sqrt{\xi_W^{}} m_W^{}\right) \right.  \nonumber \\
		&& + \frac{4 c_{\rm L}^{} c_{\rm R}^{}}{c^2} \delta_{\alpha\beta}^{} \left[(d-1) B_0^{}\left(p^2;m_\alpha^{},m_Z^{}\right) + \left(\xi_Z^{} + \frac{m_\alpha^2}{4 c_{\rm L}^{} c_{\rm R}^{} m_Z^2}\right) B_0^{}\left(p^2;m_\alpha^{},\sqrt{\xi_Z^{}} m_Z^{}\right) \right] \nonumber \\
		&& + 4 s^2 \delta_{\alpha\beta}^{} \left[(d-1)
		B_0^{}\left(p^2;m_\alpha^{},\lambda\right) + \xi_A^{} B_0^{}\left(p^2;m_\alpha^{},\sqrt{\xi_A^{}} \lambda \right) \right] \nonumber \\
		&& \left. - \delta_{\alpha\beta}^{} \frac{m_\alpha^2}{m_W^2} B_0^{}\left(p^2;m_\alpha^{},m_h^{}\right) + \delta_{\alpha\beta}^{} \frac{2 (4\pi)^2 T}{g m_h^2 m_W^{}} \vphantom{\sum_i} \right\} \;.
	\end{eqnarray}
	We introduce the fictitious mass $\lambda$ for the photon to regularize the infrared divergences, and the limit of $\lambda\to 0$ should be taken at the very end of calculations. The PV functions $B_0^{}$ and $B_1^{}$ are defined as usual.	
	
	\subsection{Self-energies of Majorana Neutrinos}
	
	\begin{figure}[t]
		\centering
		\includegraphics[scale=1]{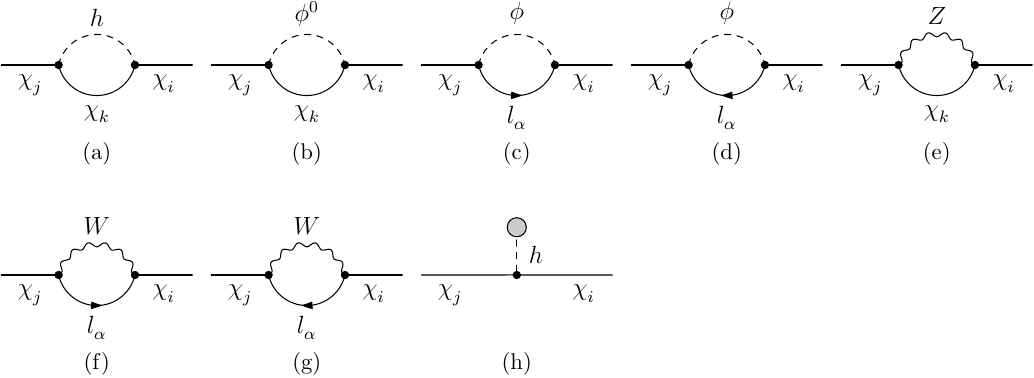}
		\vspace{-0.7cm}
		\caption{The self-energy corrections of Majorana neutrinos $\Sigma_{ij}^{} (p^2)$.}
		\label{fig:neutrino}
	\end{figure}
	
	The one-loop corrections to the self-energy of Majorana neutrinos $\chi_i^{}$ (for $i=1,\cdots,6$) are plotted in Fig.~\ref{fig:neutrino}. To calculate those diagrams, one should first determine the fermion flow and then write down the amplitudes according to the corresponding Feynman rules in Sec.~\ref{sec:typeI}. Note that there are two diagrams with $W$ or $\phi$ appearing in the loop, e.g., (c)-(d) and (f)-(g), corresponding to two different types of fermion number flow. The left-handed and scalar parts of the two-point functions are
	\begin{eqnarray}
		\Sigma_{ij}^{\rm \chi,L} &=& - \frac{g^2}{4 (4\pi)^2} \left\{ 2 \sum_\alpha {\cal B}_{\alpha i}^{} {\cal B}_{\alpha j}^{*} \frac{\widehat{m}_i^{} \widehat{m}_j^{}}{m_W^2} B_1^{}\left(p^2;m_\alpha^{},\sqrt{\xi_W^{}}m_W^{}\right) \right. \nonumber \\
		&& - 2 \sum_\alpha {\cal B}_{\alpha i}^{*} {\cal B}_{\alpha j}^{} \left[ - \frac{p^2 - m_\alpha^2}{2 p^2 m_W^2}\left(p^2 -\xi_W^{} m_W^2 + m_W^2\right) B_0^{}\left(p^2;m_W^{},\sqrt{\xi_W^{}} m_W^{}\right) \right. \nonumber \\
		&&  - \frac{p^2 - m_\alpha^2}{m_W^2}
		B_1^{}\left(p^2;m_W^{},\sqrt{\xi_W^{}} m_W^{}\right) - (d-2) B_1^{}\left(p^2;m_\alpha^{},m_W^{}\right) \nonumber \\
		&& - \frac{1}{2 p^2 m_W^2} \left(p^4 - 2 m_\alpha^2 p^2 - \xi_W^{} m_W^2 p^2 + m_\alpha^4 - m_\alpha^2 \xi_W^{} m_W^2\right) B_0^{}\left(p^2;m_\alpha^{},\sqrt{\xi_W^{}} m_W^{}\right) \nonumber \\
		&& + \frac{1}{2 p^2 m_W^2} \left(p^4 - 2 m_\alpha^2 p^2 - m_W^2 p^2 + m_\alpha^4 - m_\alpha^2 m_W^2\right) B_0^{}\left(p^2;m_\alpha^{},m_W^{}\right) \nonumber \\
		&& \left. - \frac{m_\alpha^2}{m_W^2}  B_1^{}\left(p^2;m_\alpha^{},\sqrt{\xi_W^{}} m_W^{}\right)\right] \nonumber \\
		&& + \sum_k \frac{1}{m_W^2} \left({\cal C}_{ik}^{} \widehat{m}_k^{}+{\cal C}_{ik}^{*} \widehat{m}_i^{}\right) \left({\cal C}_{kj}^{} \widehat{m}_k^{}+{\cal C}_{kj}^{*} \widehat{m}_j^{}\right) \left[B_1^{}\left(p^2;\widehat{m}_k^{},m_h^{}\right) + B_1^{}\left(p^2;\widehat{m}_k^{},\sqrt{\xi_Z^{}} m_Z^{}\right) \right] \nonumber \\
		&& + \sum_k {\cal C}_{ik}^{} {\cal C}_{kj}^{} \left[\frac{\left(p^2 - \widehat{m}_k^2\right)}{2 p^2m_W^2} \left(p^2 -\xi_Z^{} m_Z^2 + m_Z^2\right) B_0^{}\left(p^2;m_Z^{},\sqrt{\xi_Z^{}} m_Z^{} \right) \right. \nonumber \\
		&& + \frac{p^2 - \widehat{m}_k^2}{m_W^2} B_1^{}\left(p^2;m_Z^{},\sqrt{\xi_Z^{}} m_Z^{} \right) + \frac{d-2}{c^2} B_1^{} \left(p^2;\widehat{m}_k^{},m_Z^{}\right) \nonumber \\
		&& + \frac{1}{2 p^2 m_W^2} \left(p^4 - 2 \widehat{m}_k^2 p^2 - \xi_Z^{} m_Z^2 p^2 + \widehat{m}_k^4 - \widehat{m}_k^2 \xi_Z^{} m_Z^2 \right) B_0^{}\left(p^2;\widehat{m}_k^{},\sqrt{\xi_Z^{}} m_Z^{} \right) \nonumber \\
		&& \left.\left. - \frac{1}{2 p^2m_W^2} \left(p^4 - 2 \widehat{m}_k^2 p^2 - m_Z^2 p^2 + \widehat{m}_k^4 - \widehat{m}_k^2 m_Z^2 \right) B_0^{}\left(p^2;\widehat{m}_k^{},m_Z^{}\right) \right] \vphantom{\sum_i} \right\}  \;, \\
		\Sigma_{ij}^{\rm \chi,M} &=& - \frac{g^2}{8 (4\pi)^2} \left\{4 \sum_\alpha \left({\cal B}_{\alpha i}^{} {\cal B}_{\alpha j}^{*} \widehat{m}_j^{} + {\cal B}_{\alpha i}^{*}
		{\cal B}_{\alpha j}^{}  \widehat{m}_i^{} \right) \frac{m_\alpha^2}{m_W^2} B_0^{}\left(p^2;m_\alpha^{},\sqrt{\xi_W^{}} m_W^{}\right)\right. \nonumber \\
		&& - \frac{2}{c^2} \sum_k {\cal C}_{ik}^{*} {\cal C}_{kj}^{} \widehat{m}_k^{} \left[(d-1) B_0^{}\left(p^2;\widehat{m}_k^{},m_Z^{}\right) + \xi_Z^{} B_0^{}\left(p^2;\widehat{m}_k^{},\sqrt{\xi_Z^{}} m_Z^{} \right) \right] \nonumber \\
		&& - 2 \sum_k \frac{\widehat{m}_k^{} }{m_W^2} \left({\cal C}_{ik}^{} \widehat{m}_i^{}+{\cal C}_{ik}^{*} \widehat{m}_k^{}\right) \left({\cal C}_{kj}^{} \widehat{m}_k^{}+{\cal C}_{kj}^{*} \widehat{m}_j^{}\right) \left[B_0^{}\left(p^2;\widehat{m}_k^{},m_h^{}\right) - B_0^{}\left(p^2;\widehat{m}_k^{},\sqrt{\xi_Z^{}} m_Z^{}\right)\right] \nonumber \\
		&& \left. + \left({\cal C}_{ij}^{} \widehat{m}_i^{} + {\cal C}_{ij}^{*} \widehat{m}_j^{}\right) \frac{4 (4\pi)^2 T}{g m_h^2 m_W^{}} \vphantom{\sum_i} \right\} \;.
	\end{eqnarray}
	Meanwhile, we have $\Sigma_{ij}^{\rm L} = \Sigma_{ij}^{\rm R*}$ and $\Sigma_{ij}^{\rm M} = \Sigma_{ji}^{\rm M}$ for Majorana fermions.


\begin{thebibliography}{99}
		%\cite{ParticleDataGroup:2024cfk}
		\bibitem{ParticleDataGroup:2024cfk}
		S.~Navas \textit{et al.} [Particle Data Group],
		``Review of particle physics,''
		Phys. Rev. D \textbf{110}, no.3, 030001 (2024).
%		doi:10.1103/PhysRevD.110.030001
		%2192 citations counted in INSPIRE as of 28 Jul 2025
		
		%\cite{Xing:2020ijf}
		\bibitem{Xing:2020ijf}
		Z.~z.~Xing,
		``Flavor structures of charged fermions and massive neutrinos,''
		Phys. Rept. \textbf{854}, 1-147 (2020)
%		doi:10.1016/j.physrep.2020.02.001
		[arXiv:1909.09610 [hep-ph]].
		%280 citations counted in INSPIRE as of 28 Jul 2025
		
		%\cite{Minkowski:1977sc}
		\bibitem{Minkowski:1977sc}
		P.~Minkowski,
		``$\mu \to {\rm e}\gamma$ at a rate of one out of $10^{9}$ muon decays?,''
		Phys. Lett. B \textbf{67}, 421-428 (1977).
%		doi:10.1016/0370-2693(77)90435-X
		%5507 citations counted in INSPIRE as of 25 Jul 2025
		
		%\cite{Yanagida:1979as}
		\bibitem{Yanagida:1979as}
		T.~Yanagida,
		``Horizontal gauge symmetry and masses of neutrinos,''
		Conf. Proc. C \textbf{7902131}, 95-99 (1979)
		KEK-79-18-95.
		%2487 citations counted in INSPIRE as of 25 Jul 2025
		
		%\cite{Gell-Mann:1979vob}
		\bibitem{Gell-Mann:1979vob}
		M.~Gell-Mann, P.~Ramond and R.~Slansky,
		``Complex Spinors and Unified Theories,''
		Conf. Proc. C \textbf{790927}, 315-321 (1979)
		[arXiv:1306.4669 [hep-th]].
		%4194 citations counted in INSPIRE as of 25 Jul 2025
		
		%\cite{Glashow:1979nm}
		\bibitem{Glashow:1979nm}
		S.~L.~Glashow,
		``The Future of Elementary Particle Physics,''
		NATO Sci. Ser. B \textbf{61}, 687 (1980).
%		doi:10.1007/978-1-4684-7197-7{\_}15
		%865 citations counted in INSPIRE as of 25 Jul 2025
		
		%\cite{Mohapatra:1979ia}
		\bibitem{Mohapatra:1979ia}
		R.~N.~Mohapatra and G.~Senjanovi\'{c},
		``Neutrino Mass and Spontaneous Parity Nonconservation,''
		Phys. Rev. Lett. \textbf{44}, 912-915 (1980).
%		doi:10.1103/PhysRevLett.44.912
		%7063 citations counted in INSPIRE as of 25 Jul 2025
		
		%\cite{Broncano:2002rw}
		\bibitem{Broncano:2002rw}
		A.~Broncano, M.~B.~Gavela and E.~E.~Jenkins,
		``The effective Lagrangian for the seesaw model of neutrino mass and leptogenesis,''
		Phys. Lett. B \textbf{552}, 177-184 (2003)
		[erratum: Phys. Lett. B \textbf{636}, 332 (2006)]
%		doi:10.1016/S0370-2693(02)03130-1
		[arXiv:hep-ph/0210271 [hep-ph]].
		%170 citations counted in INSPIRE as of 11 Jul 2025
		
		%\cite{Broncano:2003fq}
		\bibitem{Broncano:2003fq}
		A.~Broncano, M.~B.~Gavela and E.~E.~Jenkins,
		``Neutrino physics in the seesaw model,''
		Nucl. Phys. B \textbf{672}, 163-198 (2003)
%		doi:10.1016/j.nuclphysb.2003.09.011
		[arXiv:hep-ph/0307058 [hep-ph]].
		%83 citations counted in INSPIRE as of 11 Jul 2025
		
		%\cite{Weinberg:1979sa}
		\bibitem{Weinberg:1979sa}
		S.~Weinberg,
		``Baryon- and Lepton-Nonconserving Processes,''
		Phys. Rev. Lett. \textbf{43}, 1566-1570 (1979).
%		doi:10.1103/PhysRevLett.43.1566
		%2629 citations counted in INSPIRE as of 23 Jul 2025
		
		%\cite{Chankowski:1993tx}
		\bibitem{Chankowski:1993tx}
		P.~H.~Chankowski and Z.~P\l{}uciennik,
		``Renormalization group equations for seesaw neutrino masses,''
		Phys. Lett. B \textbf{316}, 312-317 (1993)
%		doi:10.1016/0370-2693(93)90330-K
		[arXiv:hep-ph/9306333 [hep-ph]].
		%359 citations counted in INSPIRE as of 23 Jul 2025
		
		%\cite{Babu:1993qv}
		\bibitem{Babu:1993qv}
		K.~S.~Babu, C.~N.~Leung and J.~T.~Pantaleone,
		``Renormalization of the neutrino mass operator,''
		Phys. Lett. B \textbf{319}, 191-198 (1993)
%		doi:10.1016/0370-2693(93)90801-N
		[arXiv:hep-ph/9309223 [hep-ph]].
		%435 citations counted in INSPIRE as of 23 Jul 2025
		
		%\cite{King:2000hk}
		\bibitem{King:2000hk}
		S.~F.~King and N.~N.~Singh,
		``Renormalization group analysis of single right-handed neutrino dominance,''
		Nucl. Phys. B \textbf{591}, 3-25 (2000)
%		doi:10.1016/S0550-3213(00)00545-9
		[arXiv:hep-ph/0006229 [hep-ph]].
		%99 citations counted in INSPIRE as of 11 Jul 2025
		
		%\cite{Antusch:2001ck}
		\bibitem{Antusch:2001ck}
		S.~Antusch, M.~Drees, J.~Kersten, M.~Lindner and M.~Ratz,
		``Neutrino mass operator renormalization revisited,''
		Phys. Lett. B \textbf{519}, 238-242 (2001)
%		doi:10.1016/S0370-2693(01)01127-3
		[arXiv:hep-ph/0108005 [hep-ph]].
		%290 citations counted in INSPIRE as of 11 Jul 2025
		
		%\cite{Antusch:2002rr}
		\bibitem{Antusch:2002rr}
		S.~Antusch, J.~Kersten, M.~Lindner and M.~Ratz,
		``Neutrino mass matrix running for non-degenerate see-saw scales,''
		Phys. Lett. B \textbf{538}, 87-95 (2002)
%		doi:10.1016/S0370-2693(02)01960-3
		[arXiv:hep-ph/0203233 [hep-ph]].
		%154 citations counted in INSPIRE as of 11 Jul 2025
		
		%\cite{Antusch:2005gp}
		\bibitem{Antusch:2005gp}
		S.~Antusch, J.~Kersten, M.~Lindner, M.~Ratz and M.~A.~Schmidt,
		``Running neutrino mass parameters in see-saw scenarios,''
		JHEP \textbf{03}, 024 (2005)
%		doi:10.1088/1126-6708/2005/03/024
		[arXiv:hep-ph/0501272 [hep-ph]].
		%429 citations counted in INSPIRE as of 23 Jul 2025
		
		%\cite{Mei:2005qp}
		\bibitem{Mei:2005qp}
		J.~w.~Mei,
		``Running neutrino masses, leptonic mixing angles and $CP$-violating phases: From $M_Z^{}$ to $\Lambda_{\rm GUT}^{}$,''
		Phys. Rev. D \textbf{71}, 073012 (2005)
%		doi:10.1103/PhysRevD.71.073012
		[arXiv:hep-ph/0502015 [hep-ph]].
		%100 citations counted in INSPIRE as of 23 Jul 2025
		
		%\cite{Ohlsson:2013xva}
		\bibitem{Ohlsson:2013xva}
		T.~Ohlsson and S.~Zhou,
		``Renormalization group running of neutrino parameters,''
		Nature Commun. \textbf{5}, 5153 (2014)
%		doi:10.1038/ncomms6153
		[arXiv:1311.3846 [hep-ph]].
		%74 citations counted in INSPIRE as of 25 Jul 2025
		
		%\cite{Zhang:2021jdf}
		\bibitem{Zhang:2021jdf}
		D.~Zhang and S.~Zhou,
		``Complete one-loop matching of the type-I seesaw model onto the Standard Model effective field theory,''
		JHEP \textbf{09}, 163 (2021)
%		doi:10.1007/JHEP09(2021)163
		[arXiv:2107.12133 [hep-ph]].
		%51 citations counted in INSPIRE as of 11 Jul 2025
		
		%\cite{Wang:2023bdw}
		\bibitem{Wang:2023bdw}
		Y.~Wang, D.~Zhang and S.~Zhou,
		``Complete one-loop renormalization-group equations in the seesaw effective field theories,''
		JHEP \textbf{05}, 044 (2023)
%		doi:10.1007/JHEP05(2023)044
		[arXiv:2302.08140 [hep-ph]].
		%8 citations counted in INSPIRE as of 11 Jul 2025
		
		%\cite{Fukugita:1986hr}
		\bibitem{Fukugita:1986hr}
		M.~Fukugita and T.~Yanagida,
		``Baryogenesis without grand unification,''
		Phys. Lett. B \textbf{174}, 45-47 (1986).
%		doi:10.1016/0370-2693(86)91126-3
		%4608 citations counted in INSPIRE as of 28 Jul 2025
		
		%\cite{JUNO:2022mxj}
		\bibitem{JUNO:2022mxj}
		A.~Abusleme \textit{et al.} [JUNO],
		``Sub-percent precision measurement of neutrino oscillation parameters with JUNO,''
		Chin. Phys. C \textbf{46}, no.12, 123001 (2022)
%		doi:10.1088/1674-1137/ac8bc9
		[arXiv:2204.13249 [hep-ex]].
		%116 citations counted in INSPIRE as of 29 Jul 2025
		
		%\cite{Capozzi:2025wyn}
		\bibitem{Capozzi:2025wyn}
		F.~Capozzi, W.~Giar{\`e}, E.~Lisi, A.~Marrone, A.~Melchiorri and A.~Palazzo,
		``Neutrino masses and mixing: Entering the era of subpercent precision,''
		Phys. Rev. D \textbf{111}, no.9, 093006 (2025)
%		doi:10.1103/PhysRevD.111.093006
		[arXiv:2503.07752 [hep-ph]].
		%16 citations counted in INSPIRE as of 25 Jul 2025
		
		%\cite{Calibbi:2017uvl}
		\bibitem{Calibbi:2017uvl}
		L.~Calibbi and G.~Signorelli,
		``Charged Lepton Flavour Violation: An Experimental and Theoretical Introduction,''
		Riv. Nuovo Cim. \textbf{41}, no.2, 71-174 (2018)
%		doi:10.1393/ncr/i2018-10144-0
		[arXiv:1709.00294 [hep-ph]].
		%353 citations counted in INSPIRE as of 15 Dec 2025
		
		%\cite{Fernandez-Martinez:2024bxg}
		\bibitem{Fernandez-Martinez:2024bxg}
		E.~Fern{\'a}ndez-Mart{\'\i}nez, X.~Marcano and D.~Naredo-Tuero,
		``Global lepton flavour violating constraints on new physics,''
		Eur. Phys. J. C \textbf{84}, no.7, 666 (2024)
%		doi:10.1140/epjc/s10052-024-12973-6
		[arXiv:2403.09772 [hep-ph]].
		%20 citations counted in INSPIRE as of 15 Dec 2025
		
		%\cite{Rodejohann:2011mu}
		\bibitem{Rodejohann:2011mu}
		W.~Rodejohann,
		``Neutrino-less Double Beta Decay and Particle Physics,''
		Int. J. Mod. Phys. E \textbf{20}, 1833-1930 (2011)
%		doi:10.1142/S0218301311020186
		[arXiv:1106.1334 [hep-ph]].
		%643 citations counted in INSPIRE as of 15 Dec 2025
		
		%\cite{Dolinski:2019nrj}
		\bibitem{Dolinski:2019nrj}
		M.~J.~Dolinski, A.~W.~P.~Poon and W.~Rodejohann,
		``Neutrinoless Double-Beta Decay: Status and Prospects,''
		Ann. Rev. Nucl. Part. Sci. \textbf{69}, 219-251 (2019)
%		doi:10.1146/annurev-nucl-101918-023407
		[arXiv:1902.04097 [nucl-ex]].
		%612 citations counted in INSPIRE as of 15 Dec 2025
		
		%\cite{Cai:2017mow}
		\bibitem{Cai:2017mow}
		Y.~Cai, T.~Han, T.~Li and R.~Ruiz,
		``Lepton Number Violation: Seesaw Models and Their Collider Tests,''
		Front. in Phys. \textbf{6}, 40 (2018)
%		doi:10.3389/fphy.2018.00040
		[arXiv:1711.02180 [hep-ph]].
		%343 citations counted in INSPIRE as of 15 Dec 2025
		
		%\cite{Fernandez-Martinez:2023phj}
		\bibitem{Fernandez-Martinez:2023phj}
		E.~Fern{\'a}ndez-Mart{\'\i}nez, M.~Gonz{\'a}lez-L{\'o}pez, J.~Hern{\'a}ndez-Garc{\'\i}a, M.~Hostert and J.~L{\'o}pez-Pav{\'o}n,
		``Effective portals to heavy neutral leptons,''
		JHEP \textbf{09}, 001 (2023)
%		doi:10.1007/JHEP09(2023)001
		[arXiv:2304.06772 [hep-ph]].
		%70 citations counted in INSPIRE as of 15 Dec 2025
		
		%\cite{tHooft:1973mfk}
		\bibitem{tHooft:1973mfk}
		G.~'t Hooft,
		``Dimensional regularization and the renormalization group,''
		Nucl. Phys. B \textbf{61}, 455-468 (1973).
%		doi:10.1016/0550-3213(73)90376-3
		%1679 citations counted in INSPIRE as of 28 Jul 2025
		
		%\cite{Weinberg:1973xwm}
		\bibitem{Weinberg:1973xwm}
		S.~Weinberg,
		``New Approach to the Renormalization Group,''
		Phys. Rev. D \textbf{8}, 3497-3509 (1973).
%		doi:10.1103/PhysRevD.8.3497
		%810 citations counted in INSPIRE as of 21 Jul 2025
		
		%\cite{Bardeen:1978yd}
		\bibitem{Bardeen:1978yd}
		W.~A.~Bardeen, A.~J.~Buras, D.~W.~Duke and T.~Muta,
		``Deep-inelastic scattering beyond the leading order in asymptotically free gauge theories,''
		Phys. Rev. D \textbf{18}, 3998-4017 (1978)
%		doi:10.1103/PhysRevD.18.3998
		%1849 citations counted in INSPIRE as of 30 Jul 2025
		
		%\cite{Xing:2007zj}
		\bibitem{Xing:2007zj}
		Z.~z.~Xing,
		``Correlation between the charged current interactions of light and heavy Majorana neutrinos,''
		Phys. Lett. B \textbf{660}, 515-521 (2008)
%		doi:10.1016/j.physletb.2008.01.038
		[arXiv:0709.2220 [hep-ph]].
		%116 citations counted in INSPIRE as of 11 Jul 2025
		
		%\cite{Xing:2011ur}
		\bibitem{Xing:2011ur}
		Z.~z.~Xing,
		``Full parametrization of the $6 \times 6$ flavor mixing matrix in the presence of three light or heavy sterile neutrinos,''
		Phys. Rev. D \textbf{85}, 013008 (2012)
%		doi:10.1103/PhysRevD.85.013008
		[arXiv:1110.0083 [hep-ph]].
		%102 citations counted in INSPIRE as of 11 Jul 2025
		
		%\cite{Pilaftsis:1991ug}
		\bibitem{Pilaftsis:1991ug}
		A.~Pilaftsis,
		``Radiatively induced neutrino masses and large Higgs-neutrino couplings in the Standard Model with Majorana fields,''
		Z. Phys. C \textbf{55}, 275-282 (1992)
%		doi:10.1007/BF01482590
		[arXiv:hep-ph/9901206 [hep-ph]].
		%443 citations counted in INSPIRE as of 16 Jul 2025
		
		%\cite{Denner:2019vbn}
		\bibitem{Denner:2019vbn}
		A.~Denner and S.~Dittmaier,
		``Electroweak radiative corrections for collider physics,''
		Phys. Rept. \textbf{864}, 1-163 (2020)
%		doi:10.1016/j.physrep.2020.04.001
		[arXiv:1912.06823 [hep-ph]].
		%221 citations counted in INSPIRE as of 24 Jul 2025
		
		%\cite{Denner:1992vza}
		\bibitem{Denner:1992vza}
		A.~Denner, H.~Eck, O.~Hahn and J.~K\"{u}blbeck,
		``Feynman rules for fermion-number-violating interactions,''
		Nucl. Phys. B \textbf{387}, 467-481 (1992).
%		doi:10.1016/0550-3213(92)90169-C
		%343 citations counted in INSPIRE as of 11 Jul 2025
		
		%\cite{Aoki:1982ed}
		\bibitem{Aoki:1982ed}
		K.~I.~Aoki, Z.~Hioki, R.~Kawabe, M.~Konuma and T.~Muta,
		``Electroweak Theory. Framework of On-Shell Renormalization and Study of Higher-Order Effects,''
		Prog. Theor. Phys. Suppl. \textbf{73}, 1-226 (1982).
%		doi:10.1143/PTPS.73.1
		%462 citations counted in INSPIRE as of 11 Jul 2025
		
		%\cite{Bohm:1986rj}
		\bibitem{Bohm:1986rj}
		M.~B\"{o}hm, H.~Spiesberger and W.~Hollik,
		``On the 1-Loop Renormalization of the Electroweak Standard Model and its Application to Leptonic Processes,''
		Fortsch. Phys. \textbf{34}, 687-751 (1986).
%		doi:10.1002/prop.19860341102
		%596 citations counted in INSPIRE as of 29 Jul 2025
		
		%\cite{Hollik:1988ii}
		\bibitem{Hollik:1988ii}
		W.~F.~L.~Hollik,
		``Radiative Corrections in the Standard Model and Their R\^{o}le for Precision Tests of the Electroweak Theory,''
		Fortsch. Phys. \textbf{38}, 165-260 (1990).
%		doi:10.1002/prop.2190380302
		%624 citations counted in INSPIRE as of 29 Jul 2025
		
		%\cite{Denner:1991kt}
		\bibitem{Denner:1991kt}
		A.~Denner,
		``Techniques for Calculation of Electroweak Radiative Corrections at the One-Loop Level and Results for $W$-physics at LEP 200,''
		Fortsch. Phys. \textbf{41}, 307-420 (1993)
%		doi:10.1002/prop.2190410402
		[arXiv:0709.1075 [hep-ph]].
		%1204 citations counted in INSPIRE as of 21 Jul 2025
		
		%\cite{Bohm:2001yx}
		\bibitem{Bohm:2001yx}
		M.~B\"{o}hm, A.~Denner and H.~Joos,
		``Gauge Theories of the Strong and Electroweak Interaction,''
		Vieweg+Teubner Verlag Wiesbaden, 2001.
%		doi:10.1007/978-3-322-80160-9
		%78 citations counted in INSPIRE as of 11 Jul 2025

		%\cite{Sirlin:2012mh}
		\bibitem{Sirlin:2012mh}
		A.~Sirlin and A.~Ferroglia,
		``Radiative corrections in precision electroweak physics: A historical perspective,''
		Rev. Mod. Phys. \textbf{85}, no.1, 263-297 (2013)
		%		doi:10.1103/RevModPhys.85.263
		[arXiv:1210.5296 [hep-ph]].
		%74 citations counted in INSPIRE as of 29 Jul 2025
		
		%\cite{Huang:2023nqf}
		\bibitem{Huang:2023nqf}
		J.~Huang and S.~Zhou,
		``Mikheyev-Smirnov-Wolfenstein matter potential at the one-loop level in the Standard Model,''
		Phys. Rev. D \textbf{108}, no.9, 093010 (2023)
%		doi:10.1103/PhysRevD.108.093010
		[arXiv:2307.04685 [hep-ph]].
		%4 citations counted in INSPIRE as of 11 Jul 2025
		
		%\cite{Cabibbo:1963yz}
		\bibitem{Cabibbo:1963yz}
		N.~Cabibbo,
		``Unitary Symmetry and Leptonic Decays,''
		Phys. Rev. Lett. \textbf{10}, 531-533 (1963).
%		doi:10.1103/PhysRevLett.10.531
		%7996 citations counted in INSPIRE as of 28 Jul 2025
		
		%\cite{Kobayashi:1973fv}
		\bibitem{Kobayashi:1973fv}
		M.~Kobayashi and T.~Maskawa,
		``$CP$-Violation in the Renormalizable Theory of Weak Interaction,''
		Prog. Theor. Phys. \textbf{49}, 652-657 (1973).
%		doi:10.1143/PTP.49.652
		%12424 citations counted in INSPIRE as of 28 Jul 2025
		
		%\cite{Fleischer:1980ub}
		\bibitem{Fleischer:1980ub}
		J.~Fleischer and F.~Jegerlehner,
		``Radiative corrections to Higgs-boson decays in the Weinberg-Salam Model,''
		Phys. Rev. D \textbf{23}, 2001-2026 (1981).
%		doi:10.1103/PhysRevD.23.2001
		%375 citations counted in INSPIRE as of 11 Jul 2025
		
		%\cite{Bollini:1972ui}
		\bibitem{Bollini:1972ui}
		C.~G.~Bollini and J.~J.~Giambiagi,
		``Dimensional Renormalization: The Number of Dimensions as a Regularizing Parameter,''
		Nuovo Cim. B \textbf{12}, 20-26 (1972).
%		doi:10.1007/BF02895558
		%1135 citations counted in INSPIRE as of 28 Jul 2025
		
		%\cite{tHooft:1972tcz}
		\bibitem{tHooft:1972tcz}
		G.~'t Hooft and M.~J.~G.~Veltman,
		``Regularization and renormalization of gauge fields,''
		Nucl. Phys. B \textbf{44}, 189-213 (1972).
%		doi:10.1016/0550-3213(72)90279-9
		%5569 citations counted in INSPIRE as of 28 Jul 2025
		
		%\cite{Marciano:1980pb}
		\bibitem{Marciano:1980pb}
		W.~J.~Marciano and A.~Sirlin,
		``Radiative corrections to neutrino-induced neutral-current phenomena in the ${\rm SU}(2)_L \times {\rm U}(1)$ theory,''
		Phys. Rev. D \textbf{22}, 2695 (1980)
		[erratum: Phys. Rev. D \textbf{31}, 213 (1985)].
%		doi:10.1103/PhysRevD.22.2695
		%995 citations counted in INSPIRE as of 23 Jul 2025
		
		%\cite{Degrassi:1992ff}
		\bibitem{Degrassi:1992ff}
		G.~Degrassi and A.~Sirlin,
		``Gauge dependence of basic electroweak corrections of the Standard Model,''
		Nucl. Phys. B \textbf{383}, 73-92 (1992).
%		doi:10.1016/0550-3213(92)90671-W
		%96 citations counted in INSPIRE as of 11 Jul 2025
		
		%\cite{Kniehl:1996bd}
		\bibitem{Kniehl:1996bd}
		B.~A.~Kniehl and A.~Pilaftsis,
		``Mixing renormalization in Majorana neutrino theories,''
		Nucl. Phys. B \textbf{474}, 286-308 (1996)
%		doi:10.1016/0550-3213(96)00280-5
		[arXiv:hep-ph/9601390 [hep-ph]].
		%109 citations counted in INSPIRE as of 11 Jul 2025
		
		%\cite{Grimus:2016hmw}
		\bibitem{Grimus:2016hmw}
		W.~Grimus and M.~L{\"o}schner,
		``Revisiting on-shell renormalization conditions in theories with flavor mixing,''
		Int. J. Mod. Phys. A \textbf{31}, no.24, 1630038 (2016)
		[erratum: Int. J. Mod. Phys. A \textbf{32}, no.13, 1792001 (2017)]
%		doi:10.1142/S0217751X16300386
		[arXiv:1606.06191 [hep-ph]].
		%10 citations counted in INSPIRE as of 11 Jul 2025
		
		%\cite{Denner:1990yz}
		\bibitem{Denner:1990yz}
		A.~Denner and T.~Sack,
		``Renormalization of the quark mixing matrix,''
		Nucl. Phys. B \textbf{347}, 203-216 (1990).
%		doi:10.1016/0550-3213(90)90557-T
		%131 citations counted in INSPIRE as of 17 Jul 2025
		
		%\cite{Passarino:1978jh}
		\bibitem{Passarino:1978jh}
		G.~Passarino and M.~J.~G.~Veltman,
		``One-loop corrections for ${\rm e}^+ {\rm e}^-$ annihilation into $\mu^+ \mu^-$ in the Weinberg model,''
		Nucl. Phys. B \textbf{160}, 151-207 (1979).
%		doi:10.1016/0550-3213(79)90234-7
		%3070 citations counted in INSPIRE as of 28 Jul 2025
		
		%\cite{Pilaftsis:2002nc}
		\bibitem{Pilaftsis:2002nc}
		A.~Pilaftsis,
		``Gauge and scheme dependence of mixing matrix renormalization,''
		Phys. Rev. D \textbf{65}, 115013 (2002)
%		doi:10.1103/PhysRevD.65.115013
		[arXiv:hep-ph/0203210 [hep-ph]].
		%53 citations counted in INSPIRE as of 11 Jul 2025
		
		%\cite{Nakano:1953zz}
		\bibitem{Nakano:1953zz}
		T.~Nakano and K.~Nishijima,
		``Charge Independence for $V$-particles,''
		Prog. Theor. Phys. \textbf{10}, 581-582 (1953).
%		doi:10.1143/PTP.10.581
		%415 citations counted in INSPIRE as of 16 Jul 2025
		
		%\cite{Gell-Mann:1956iqa}
		\bibitem{Gell-Mann:1956iqa}
		M.~Gell-Mann,
		``The Interpretation of the New Particles as Displaced Charge Multiplets,''
		Nuovo Cim. \textbf{4}, no.S2, 848-866 (1956).
%		doi:10.1007/BF02748000
		%236 citations counted in INSPIRE as of 16 Jul 2025
		
		%\cite{Pontecorvo:1957cp}
		\bibitem{Pontecorvo:1957cp}
		B.~Pontecorvo,
		``Mesonium and Antimesonium,''
		Sov. Phys. JETP \textbf{6}, 429-431 (1958).
		%2547 citations counted in INSPIRE as of 23 Jul 2025
		
		%\cite{Maki:1962mu}
		\bibitem{Maki:1962mu}
		Z.~Maki, M.~Nakagawa and S.~Sakata,
		``Remarks on the Unified Model of Elementary Particles,''
		Prog. Theor. Phys. \textbf{28}, 870-880 (1962).
%		doi:10.1143/PTP.28.870
		%5400 citations counted in INSPIRE as of 25 Jul 2025
		
		%\cite{Pontecorvo:1967fh}
		\bibitem{Pontecorvo:1967fh}
		B.~Pontecorvo,
		``Neutrino Experiments and the Problem of Conservation of Leptonic Charge,''
		Zh. Eksp. Teor. Fiz. \textbf{53}, 1717-1725 (1967).
		%2716 citations counted in INSPIRE as of 22 Jul 2025
		
		%\cite{Xing:2023kdj}
		\bibitem{Xing:2023kdj}
		Z.~z.~Xing,
		``CP violation in light neutrino oscillations and heavy neutrino decays: A general and explicit seesaw-bridged correlation,''
		Phys. Lett. B \textbf{844}, 138065 (2023)
%		doi:10.1016/j.physletb.2023.138065
		[arXiv:2306.02362 [hep-ph]].
		%5 citations counted in INSPIRE as of 23 Jul 2025
		
		%\cite{Xing:2024xwb}
		\bibitem{Xing:2024xwb}
		Z.~z.~Xing,
		``Mapping the sources of CP violation in neutrino oscillations from the seesaw mechanism,''
		Phys. Lett. B \textbf{856}, 138909 (2024)
%		doi:10.1016/j.physletb.2024.138909
		[arXiv:2406.01142 [hep-ph]].
		%7 citations counted in INSPIRE as of 23 Jul 2025
		
		%\cite{Xing:2024gmy}
		\bibitem{Xing:2024gmy}
		Z.~z.~Xing and J.~y.~Zhu,
		``Confronting the seesaw mechanism with neutrino oscillations: A general and explicit analytical bridge,''
		Nucl. Phys. B \textbf{1018}, 117041 (2025)
%		doi:10.1016/j.nuclphysb.2025.117041
		[arXiv:2412.17698 [hep-ph]].
		%4 citations counted in INSPIRE as of 29 Jul 2025
		
		%\cite{Alam:2022cdv}
		\bibitem{Alam:2022cdv}
		Z.~Alam and S.~P.~Martin,
		``Standard model at 200~GeV,''
		Phys. Rev. D \textbf{107}, no.1, 013010 (2023)
%		doi:10.1103/PhysRevD.107.013010
		[arXiv:2211.08576 [hep-ph]].
		%11 citations counted in INSPIRE as of 23 Jul 2025
		
		%\cite{Hahn:2000kx}
		\bibitem{Hahn:2000kx}
		T.~Hahn,
		``Generating Feynman diagrams and amplitudes with {\it FeynArts} 3,''
		Comput. Phys. Commun. \textbf{140}, 418-431 (2001)
%		doi:10.1016/S0010-4655(01)00290-9
		[arXiv:hep-ph/0012260 [hep-ph]].
		%2320 citations counted in INSPIRE as of 28 Jul 2025
		
		%\cite{Patel:2015tea}
		\bibitem{Patel:2015tea}
		H.~H.~Patel,
		``{\it Package}-X: A {\it Mathematica} package for the analytic calculation of one-loop integrals,''
		Comput. Phys. Commun. \textbf{197}, 276-290 (2015)
%		doi:10.1016/j.cpc.2015.08.017
		[arXiv:1503.01469 [hep-ph]].
		%574 citations counted in INSPIRE as of 25 Jul 2025
		
		%\cite{Patel:2016fam}
		\bibitem{Patel:2016fam}
		H.~H.~Patel,
		``{\it Package}-X 2.0: A {\it Mathematica} package for the analytic calculation of one-loop integrals,''
		Comput. Phys. Commun. \textbf{218}, 66-70 (2017)
%		doi:10.1016/j.cpc.2017.04.015
		[arXiv:1612.00009 [hep-ph]].
		%302 citations counted in INSPIRE as of 24 Jul 2025
	\end{thebibliography}
\end{document}